\documentclass[a4paper]{article}

\usepackage{jcappub}
\usepackage{amsmath}
\usepackage{amssymb}
\usepackage{subcaption}
\usepackage{glossaries}
\usepackage{multirow}
\usepackage{xcolor}

\makeglossaries
\newacronym{sbgw}{SBGW}{Stochastic Background of Gravitational Waves}
\newacronym{gw}{GW}{Gravitational Wave}
\newacronym{slc}{SLC}{Small loop contribution}
\newacronym{llc}{LLC}{Large loop contribution}
\newacronym{lnd}{LND}{Loop Number Density}
\newacronym{gr}{GR}{General Relativity}
\newacronym{slnd}{SLND}{Standard Loop Number Density}
\newacronym{epsl}{EPSL}{Extra Population of Small Loops}
\newacronym{lrs}{LRS}{Lorenz, Ringeval and Sakellariadou}
\newacronym{bos}{BOS}{Blanco-Pillado, Olum and Shlaer}

\newlength{\roww}
\newcommand{\ugw}{\mathrm{gw}}
\newcommand{\ur}{\mathrm{r}}
\newcommand{\um}{\mathrm{m}}
\newcommand{\uc}{\mathrm{c}}
\newcommand{\ud}{\mathrm{d}}
\newcommand{\uh}{\mathrm{h}}
\newcommand{\uH}{\mathrm{H}}

\newcommand{\uir}{\mathrm{ir}}
\newcommand{\udh}{\ud_\uh}
\newcommand{\uHr}{\uH_\ur}
\newcommand{\uHm}{\uH_\um}
\newcommand{\uHo}{\uH_0}
\newcommand{\ucrit}{\mathrm{crit}}
\newcommand{\OmegaGW}{\Omega_{\ugw}}
\newcommand{\OmegaR}{\Omega_\ur}
\newcommand{\OmegaM}{\Omega_\um}
\newcommand{\rhoGW}{\rho_{\ugw}}
\newcommand{\rhoi}{\rho_{\infty}}
\newcommand{\calF}{\mathcal{F}}
\newcommand{\C}{\mathcal{C}}
\newcommand{\calP}{\mathcal{P}}
\newcommand{\fa}{f_\mathrm{a}}
\newcommand{\fb}{f_\mathrm{b}}
\newcommand{\fc}{f_\mathrm{c}}
\newcommand{\fd}{f_\mathrm{d}}
\newcommand{\fone}{f_\mathrm{1}}
\newcommand{\ftwo}{f_\mathrm{2}}
\newcommand{\fthre}{f_\mathrm{3}}
\newcommand{\ffour}{f_\mathrm{4}}
\newcommand{\Pm}{\mathrm{P}_\um}
\newcommand{\Q}{\mathrm{Q}}
\newcommand{\Qr}{\mathrm{Q}_\ur}
\newcommand{\Qm}{\mathrm{Q}_\um}
\newcommand{\Qrm}{\mathrm{Q}_{\ur\um}}

\newcommand{\tQr}{\tilde{\mathrm{Q}}_\ur}
\newcommand{\tQm}{\tilde{\mathrm{Q}}_\um}
\newcommand{\tQrm}{\tilde{\mathrm{Q}}_{\ur\um}}

\newcommand{\hQr}{\hat{\mathrm{Q}}_\ur}
\newcommand{\hQm}{\hat{\mathrm{Q}}_\um}

\newcommand{\gammad}{\gamma_\ud}
\newcommand{\gammac}{\gamma_\uc}

\newcommand{\gammai}{\gamma_\infty}
\newcommand{\gammair}{\gamma_{\uir}}
\newcommand{\chis}{\chi_*}
\newcommand{\chicrit}{\chi_{_\ucrit}}
\newcommand{\chir}{\chi_{_\ur}}
\newcommand{\chim}{\chi_{_\um}}
\newcommand{\chiir}{\chi_{\uir}}
\newcommand{\meps}{\epsilon}
\newcommand{\mur}{\meps_{_\ur}}
\newcommand{\mum}{\meps_{_\um}}
\newcommand{\ucr}{c_{_\ur}}
\newcommand{\ucm}{c_{_\um}}
\newcommand{\zeq}{{z_\mathrm{eq}}}
\newcommand{\teq}{{t_\mathrm{eq}}}
\newcommand{\hypergauss}[4]{\phantom{}_{_2}\mathrm{F}\!_{_1}\!\left(#1,#2;#3;#4\right)}

\title{Impact of the small-scale structure on the Stochastic Background of Gravitational Waves from cosmic strings}
\author[a]{Pierre Auclair}
\affiliation[a]{Universit\'e de Paris, CNRS, Astroparticule et Cosmologie, F-75013 Paris, France }
\emailAdd{auclair@apc.in2p3.fr}
\date{\today}

\begin{document}

\abstract{Numerical simulations and analytical models suggest that infinite cosmic strings produce cosmic string loops of all sizes with a given power-law.
Precise estimations of the power-law exponent are still matter of debate while numerical simulations do not incorporate all the radiation and back-reaction effects expected to affect the network at small scales.
Previously it has been shown, using a Boltzmann approach, that depending on the steepness of the loop production function and the gravitational back-reaction scale, a so-called \gls{epsl} can be generated in the loop number density.
We propose a framework to study the influence of this extra population of small loops on the \gls{sbgw}.
We show that this extra population can have a significant signature at frequencies higher than $\uHo(\Gamma G\mu)^{-1}$ where $\Gamma$ is of order $50$ and $\uHo$ is the Hubble constant.
We propose a complete classification of the \gls{gw} power spectra expected from cosmic strings into four classes, including the model of Blanco-Pillado, Olum and Shlaer and the model of Lorenz, Ringeval and Sakellariadou.
Finally we show that given the uncertainties on the Polchinski-Rocha exponents, two hybrid classes of \gls{gw} power spectrum can be considered giving very different predictions for the \gls{sbgw}.
}

\maketitle
\section{Introduction}

    The first direct observation of \acrfullpl{gw} coming from the merger of two black holes \cite{abbott_observation_2016} was both a wonderful check of the theory of General Relativity and the onset of \gls{gw} astronomy.
    Since \gls{gw} propagate freely throughout the Universe, they are not limited by the last scattering surface, and give us an unprecedented opportunity to look for topological defects, and in particular cosmic strings.

    Cosmic strings are one-dimensional topological defects that may have formed during a symmetry-breaking phase transition in the early Universe \cite{kibble_topology_1976,hindmarsh_cosmic_1995,vilenkin_cosmic_2001,vachaspati_cosmic_2015}.
    Nambu-Goto strings are a powerful one-dimensional approximation to study these solitonic solutions on cosmological scales.
    The evolution of a Nambu-Goto string network in an expanding background has been studied both analytically  \cite{kibble_topology_1976,vachaspati_formation_1984,kibble_evolution_1985,allen_small-scale_1991,allen_kinky_1991,austin_evolution_1993,polchinski_analytic_2006,polchinski_cosmic_2007,polchinski_cosmic_2007-1,lorenz_cosmic_2010,blanco-pillado_number_2014,martins_models_2014,vieira_models_2016,auclair_cosmic_2019} and through numerical simulations \cite{bennett_cosmic-string_1989,albrecht_evolution_1989,bennett_high-resolution_1990,allen_cosmic-string_1990,blanco-pillado_large_2011,martins_fractal_2006,ringeval_cosmological_2007} in the last decades, and is still subject of intense research.

    A general result is that the network relaxes to an attractor solution known as the scaling solution and remains self-similar with the Hubble radius.
    If cosmic strings were formed, scaling means they survive during the whole history of the Universe and are present all over the sky.
    Strings can induce anisotropies on the Cosmic Microwave Background and have been searched for in the Planck data .
    The current CMB constraints give an upper bound for the string tension $\mu$ of $G\mu < 1.5 \times 10^{-7}$ for Nambu-Goto strings and $G\mu < 2 \times 10^{-7}$ for Abelian-Higgs strings, where $G$ is Newton's constant \cite{jeong_search_2005,ringeval_cosmic_2010,planck_collaboration_planck_2014, Lizarraga:2016onn}.

    These bounds are calculated assuming a given scenario for the evolution of the loop number density throughout the history of the Universe (see below), and can depend a lot on those assumptions.
    Furthermore, each closed cosmic string loop radiates \gls{gw} and the superposition of them produces a \acrfull{sbgw}  \cite{vachaspati_gravitational_1985,hindmarsh_gravitational_1990,allen_gravitational_1992,damour_gravitational_2001,siemens_gravitational_2001,siemens_gravitational_2006} which could be detected by gravitational wave detectors.
    This background has been looked for in LIGO/Virgo for O1 and O2 and gives already a tighter upper bound for $G\mu$ which is, however, very dependent on the cosmic string model used, ranging from $G\mu < 1.1\times 10^{-6}$ to $G\mu < 2.1 \times 10^{-14}$ \cite{abbott_constraints_2018,the_ligo_scientific_collaboration_search_2019}.
    In section \ref{sec:gwdetect} we will explain the origin of the orders of magnitude difference between these two constraints.
    The most stringent and stable constraint today comes from pulsar timing experiment giving $G\mu \lesssim 10^{-10}$ \cite{sanidas_constraints_2012}.

    Building a model for the evolution of the cosmic string network is challenging, and involves both analytical modelling and numerical simulations.
    Nambu-Goto simulations are necessary to determine the large-scale behavior of the loop number density, but are unable to provide a description of the smallest scales as they do not include gravitational radiation nor the back-reaction that dominates on these scales \cite{blanco-pillado_gravitational_2018,Blanco-Pillado:2019nto,chernoff_gravitational_2019}.
    One of the difficulties is the proliferation of kinks -- which are discontinuities in the tangent vector of the string.
    Kinks are formed every time two strings intersect each other, are removed by outgoing loops and are smoothed by gravitational back-reaction.
    If the scaling of the large scales is today well supported by numerical simulations, the build-up of a population of kinks has raised some doubts on the scaling properties of the small-scales \cite{polchinski_analytic_2006,austin_evolution_1993,copeland_kinks_2009,allen_small-scale_1991,allen_kinky_1991,vieira_models_2016,martins_models_2014,siemens_size_2002,martins_fractal_2006} and this situation cannot be settled with simulations available today.
    A first attempt to model analytically the number of kinks using the one-scale model was performed in  \cite{allen_small-scale_1991,allen_kinky_1991}, and showed that kinks accumulate until the number of kinks reaches a scaling regime introducing another scale to the system \cite{copeland_kinks_2009}.
    Models were later introduced to take into account this small-scale structure, these include the three-scale model \cite{austin_evolution_1993}, a renormalized velocity-dependent one-scale model \cite{vieira_models_2016,martins_models_2014} and the Polchinski-Rocha model based on fractal dimensions \cite{polchinski_analytic_2006,dubath_cosmic_2008,siemens_gravitational_2001,polchinski_cosmic_2007} which we will use in the following.
    It introduces a positive exponent $\chi$ defined later in the equation \eqref{eq:looprod}, and one of its particular prediction is that the gravitational back-reaction scale is not $\Gamma G\mu t$ as in \cite{siemens_gravitational_2001,siemens_size_2002}, but rather the smaller scale $\Upsilon (G\mu)^{1+2\chi} t$ where $\Upsilon$ is of order 20.

    The goal of this article is to provide a unified framework which can continuously describe, with a limited set of parameters, different cosmic string loop models from the literature and give predictions for the \gls{sbgw}.
    It is built using the analytical model of Polchinski and Rocha \cite{polchinski_analytic_2006,polchinski_cosmic_2007,polchinski_cosmic_2007-1} and later developments \cite{auclair_cosmic_2019,lorenz_cosmic_2010}, and therefore includes the parameter $\chi$.
   With this framework, we aim at gaining a deeper understanding of the \gls{sbgw} and why constraints on the string tension from LIGO/Virgo are so model-dependent.
    We also expect to use this framework to give model-independent constraints on the string tension.

   Using our unified framework, we can furthermore focus on two particular models, the \gls{bos} \cite{blanco-pillado_number_2014} and the \gls{lrs} \cite{lorenz_cosmic_2010} models.
    The \gls{bos} model is based on the simulations conducted in \cite{blanco-pillado_number_2014,blanco-pillado_large_2011} and makes the assumption that the production of loops with sizes smaller than the gravitational radiation scale $t \Gamma G\mu$, where $\Gamma \approx 50$, is suppressed.
    On the other hand, the \gls{lrs} model is based on the simulations conducted in \cite{ringeval_cosmological_2007} and based on the analytical studies of \cite{polchinski_analytic_2006,polchinski_cosmic_2007} which assume that small loops are produced down to the gravitational back-reaction scale, which is smaller than the gravitational radiation scale by several orders of magnitude.
    As a result the two models give very different predictions for the loop number density.
    Relative to the first one, the second gives rise to an \acrfull{epsl}.
    The smaller back-reaction scale \emph{\`a la} Polchinski-Rocha can be introduced in the \gls{bos} model producing also an additional population of small loops \cite{auclair_cosmic_2019}.
    It is therefore interesting to understand its effect on the \gls{sbgw}.

    This paper is set up as follows.
    Section \ref{sec:framework} describes the theoretical framework used to unify several cosmic string models found in the literature.
    In particular, we show that the loop number density is naturally composed of two distinct population, a \gls{slnd} which is very similar to the prediction of the one-scale model, and an \gls{epsl}.
    Section \ref{sec:sbgw} shows how to calculate analytically an estimate to the \gls{sbgw} from cosmic strings and discusses the validity of the approximations made.
    Section \ref{sec:results} then combines the results to obtain the dependence of several types of \gls{gw} experiments to the uncertainties on the cosmic string parameters.
    Finally, section \ref{sec:conclusion} presents our conclusions.

\section{Theoretical framework}
\label{sec:framework}
    \subsection{The network of infinite strings}
    \label{sec:infinite}
        A standard way to model the evolution of cosmic strings is to study \emph{infinite strings} and \emph{closed loops} as two distinct populations in interaction.
        These infinite strings of cosmological sizes are stretched by the expansion of the universe characterized by the scale factor $a(t)$ which evolves as $t^\nu$ where $\nu=1/2$ in the radiation-dominated era and $\nu=2/3$ in the matter-dominated era.
        At the same time, they lose energy by forming loops.
       Closed loops are formed when two infinite strings intersect each other or when one self-intersects.
        In principle these loops can rejoin the infinite strings
        or fragment into smaller loops.
        At the end of the fragmentation, one is left with with smaller non-self intersecting long-lived loops.
        It is this population of long-lived non self-intersecting loops that dominates the \gls{sbgw} and that we model.

        In this article, we assume the inter-commutation probability to be equal to one, although some types of cosmic strings may have it strictly smaller than one \cite{vilenkin_string-dominated_1984}.
        Based on analytical models \cite{vilenkin_cosmic_2001} and numerical simulations \cite{ringeval_cosmological_2007,blanco-pillado_number_2014,blanco-pillado_large_2011} we expect the network of infinite strings to \emph{scale} in radiation-dominated or in matter-dominated era.
        Scaling is an attractor solution of the network in which all the relevant length scales are proportional to the horizon size $\udh$ which itself is proportional to the cosmic time $t$.
        During scaling the energy density contained in cosmic strings evolves as $\rhoi \propto t^{-2}$.

        The loop production function $\calP(\ell,t)$ is the number of long-lived non self-intersecting loops of invariant length $\ell$ per unit volume per unit time formed at cosmic time $t$.
        In scaling, $t^5 \calP(\ell,t)$ is expected to be only a function of the scaling variable $\gamma = \ell/t$.
        There exist different calculations in the literature concerning the shape of this loop production function.
        In the one-scale model introduced in \cite{kibble_evolution_1985}, all loops are assumed to be formed with the same size, meaning the loop production function is a Dirac-delta distribution.
        This typical size is then inferred from numerical simulations.
        In the work of \cite{polchinski_analytic_2006,polchinski_cosmic_2007}, it has been argued that the loop production function is a power-law, something which was found in the simulations of \cite{martins_fractal_2006} and is compatible with the simulations of \cite{ringeval_cosmological_2007}.
        In such a case the loop production function is parameterized by a parameter $\chi$ and a multiplicative constant $c$
        \begin{equation}
            t^5 \calP(\ell,t) = c \gamma^{2\chi-3} \text{ for } \gammac < \gamma
            \label{eq:looprod}
        \end{equation}
        where the analytical study of the small-scale structure of \cite{polchinski_analytic_2006} suggested the introduction of a \emph{gravitational back-reaction scale} $\gammac$ below which the production of loops by the network is suppressed
        \begin{equation}
            \gammac = \Upsilon (G\mu)^{1+2\chi}
        \end{equation}
        where $\Upsilon$ is of order $20$.
        This loop production function was developed in an attempt to take into account the small-scale structure of the network.
        It was shown in \cite{lorenz_cosmic_2010} that the precise shape of the loop production function below $\gammac$ has only a small impact on the \gls{lnd}.
        It has been used in \cite{ringeval_stochastic_2017,lorenz_cosmic_2010} to calculate the loop number density and leads to a significant \acrfull{epsl} with respect to the one-scale scenario.
        To fit the numerical simulations of \cite{ringeval_cosmological_2007}, their analysis assumed the network to be \emph{sub-critical} meaning $\chi < \chicrit$ where
        \begin{equation}
            \chicrit = \dfrac{3\nu-1}{2} = \left\{
            \begin{array}{ll}
                1/4 & \text{in radiation era} \\
                1/2 & \text{in matter era}
            \end{array}\right.
        \end{equation}
        Critical $\chi=\chicrit$ and super-critical $\chi > \chicrit$ networks were finally studied in \cite{auclair_cosmic_2019}.
        This super-critical regime is supported by the Nambu-Goto simulations of \cite{blanco-pillado_number_2014}.
        It is therefore important to include super-critical regimes in our framework for future applications.

    \subsection{Loop number density}

        Once the loop production function is known, it can be injected into the Boltzmann equation for the \gls{lnd} $\calF(\ell,t)$
        \begin{equation}
            \label{eq:dpart}
            \dfrac{\ud}{\ud t} \left(a^3\calF \right) = a^3 \calP(\ell,t),
        \end{equation}
        where the effect of the expansion of the universe is taken into account by introducing the scale factor $a$.
        The loops radiate \gls{gw} with a rate we assume to be constant and given by \cite{blanco-pillado_stochastic_2017,allen_gravitational_1992}
        \begin{equation}
            \dfrac{\ud \ell}{\ud t} = -\Gamma G\mu \equiv -\gammad
        \end{equation}
        where $\Gamma$ is of order 50 \cite{vachaspati_gravitational_1985, blanco-pillado_stochastic_2017}.
        This Boltzmann equation can be solved if one assumes either radiation or matter domination and that the network of infinite strings is scaling so that the loop production function scales and is given by equation \eqref{eq:looprod}.
        The complete set of solutions can be found in \cite{auclair_cosmic_2019}.
        The loop number density no longer necessarily scales, unless one assumes that the loop production function is cutoff for $\gamma \ge \gammai$, where $\gammai$ is expected to be of the order of the Hubble horizon.
        The authors suggest the inclusion of a sharp infrared cutoff to regularize those new solutions and showed that the he precise shape of the cutoff only has a small effect on the loop distribution. We neglect it in the remainder of this paper.
        Even in these critical and super-critical regimes, one can observe a large population of small loops in the \gls{lnd} up to a new value of $\chi=\chiir$
        \begin{equation}
        \chiir = \dfrac{1+\sqrt{12\nu-3}}{4} = \begin{cases}
            \approx 0.68 \text{ in radiation era} \\
            \approx 0.8 \text{ in matter era}
        \end{cases}> \chicrit
        \end{equation}
        introducing an additional knee in the \gls{lnd} at
        \begin{equation}
            \gammair = \left(\dfrac{- \meps \gammai^{\meps}}{2-2\chi}\right)^{1/(2-2\chi)}\gammad^{(3-3\nu)/(2-2\chi)}
        \end{equation}
        in which $\meps$ is given by \footnote{This parameter is noted $\mu$ in \cite{auclair_cosmic_2019,lorenz_cosmic_2010,ringeval_stochastic_2017}. We change the notation to avoid confusion with the string tension.}
        \begin{equation}
            \meps \equiv 3 \nu - 2\chi -1
            \label{eq:meps}
        \end{equation}
        The fact that critical and super-critical models present an extra population of small loops motivates us to study the impact of this population on the \gls{sbgw}


    \subsection{Normalization of the loop production function}
    \label{sec:norm}

    \begin{figure}
        \centering
        \begin{subfigure}{0.49\textwidth}
            \includegraphics[width=\textwidth]{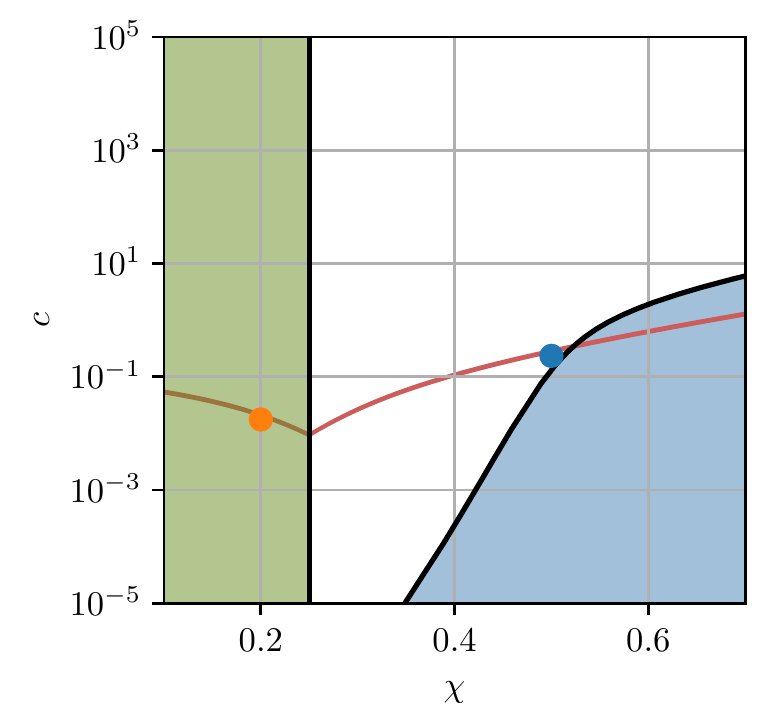}
            \caption{Radiation era}
        \end{subfigure}
        \begin{subfigure}{0.49\textwidth}
            \includegraphics[width=\textwidth]{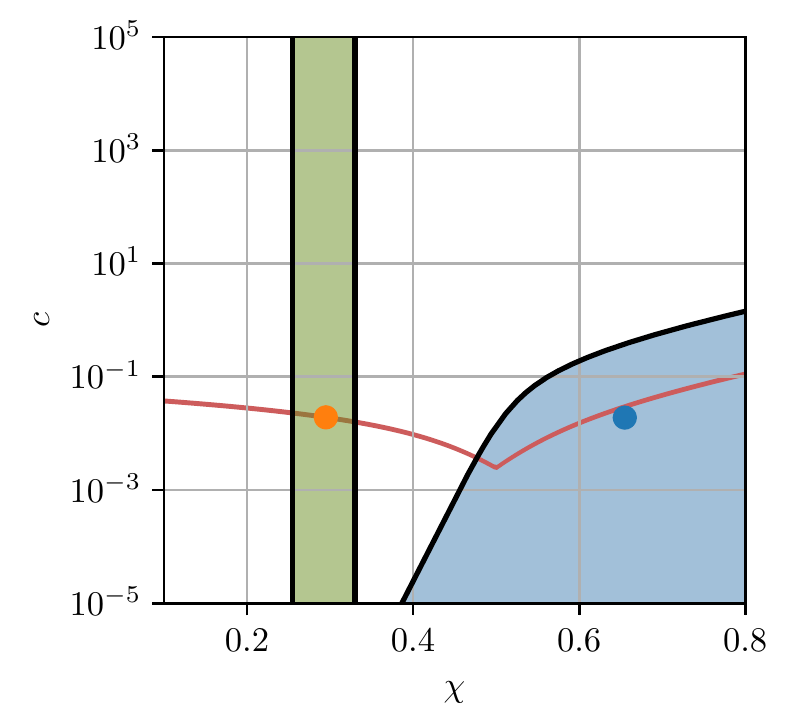}
            \caption{Matter era}
        \end{subfigure}
        \caption{Normalization of the loop production function. The boundary of the blue region is given by the "one-scale energy balance". The green region is given by measurements in \cite{ringeval_cosmological_2007}.
        The red line shows the set of parameters giving order unity loops per Hubble radius, see section \ref{sec:decompose}.
        The blue dot corresponds to the parameters of the \gls{bos} model, and the orange dot to those of the \gls{lrs} model.}
        \label{fig:energy-balance}
    \end{figure}

    Currently there is a debate on how to normalize the loop production function, that is the constant $c$ in (2.1) based on measurements from numerical simulations.
    In this section, we will review two different approaches followed in the community.

    The first approach --- explicitly stated in \cite{blanco-pillado_large_2011, Blanco-Pillado:2019vcs}, and implicitly used in the one-scale model \cite{vilenkin_cosmic_2001} --- is to use an energy conservation equation to put an upper bound on the energy lost by the network of infinite strings into loops.
    Assuming that the energy density of the infinite string network $\rhoi$ is lost through the expansion of the Universe, redshifting and by the formation of non-self-intersecting loops \cite{vilenkin_cosmic_2001}
    \begin{equation}
        \dfrac{\ud \rhoi}{\ud t}= - 2 \uH (1+\langle v_\infty^2 \rangle)\rhoi - \mu \int \ell \calP(\ell,t) \ud \ell
        \label{eq:energy-balance}
    \end{equation}
    where $\uH$ is the Hubble parameter and $\langle v_\infty^2 \rangle$ is the average velocity of the infinite strings and has been measured to be $0.45$ (resp. $0.40$,$0.35$) in a flat space-time (resp. radiation dominated, matter-dominated) \cite{blanco-pillado_large_2011}.
    On assuming that the scale factor $a\propto t^\nu$ and inserting \eqref{eq:looprod} into \eqref{eq:energy-balance}, the energy density of the infinite strings
    \begin{equation}
        \rhoi(t) = \dfrac{c \mu t^{-2}}{2\left[1-\nu(1+\langle v_\infty^2\rangle)\right]} \int_{\gammac}^{\gammai} \gamma^{2\chi-2} \ud \gamma
        \label{eq:rhoi}
    \end{equation}
    is the well known attractor scaling solution $\rhoi \propto \mu t^{-2}$.
    This can be compared to the values found for each era in numerical simulations and used to give an upper bound for the parameter $c$ once $\chi$, $\gammac$ and $\gammai$ are fixed.
    The corresponding allowed parameter space for $(c,\chi)$ is denoted as "one-scale energy balance" in figure \ref{fig:energy-balance}.
    It should be noted that numerical simulations do not include any gravitational radiation nor back-reaction, meaning that there the only equivalent to a lower cutoff in the integral of equation \eqref{eq:rhoi} is determined by the smallest length-scale set at the initialization of the simulation.
    If $\chi \leq 1/2$, the integral is dominated by this nonphysical lower bound, and one expect $t^2 \rhoi$ to diverge if the simulation is long enough \cite{Blanco-Pillado:2019vcs}.

    Another approach advocated in \cite{lorenz_cosmic_2010} is to consider only the large scale \gls{lnd} determined in simulations as trustworthy.
    It can be parameterized as a power-law on large-scales $\gamma > \gammad$ and fitted to the analytical predictions \cite{auclair_cosmic_2019}
    \begin{equation}
        t^4 \calF = A \gamma^{-p}.
    \end{equation}
    In the numerical simulations of \cite{ringeval_cosmological_2007}, they obtain a value of $A$ which is compatible with other numerical simulations.
    As shown in section \ref{sec:decompose}, the value of $A$ is related to the parameters of the loop production function $(c, \chi)$.
    Hence a given value of $A$ determines a curve in the $(c,\chi)$ which is the red line of figure \ref{fig:energy-balance}.

    While there seems to be a general agreement for the parameter $A$, there is a strong tension on the parameter $p$.
    Even though the uncertainty interval given for $p$ in \cite{ringeval_cosmological_2007} does not exclude the degenerate value $5/2$ in the radiation era, their best fit systematically points to an higher value than 5/2 and the authors have used the best fit value $p=2.6$ since then, thus selecting the green region of parameter space denoted in figure \ref{fig:energy-balance}.

    One can see that these two interpretations of two different numerical simulations do not agree on the values for the different parameters.
    It should be noted that the loop production function has been measured directly in \cite{blanco-pillado_large_2011} giving values for $(c,\chi)$ compatible with the energy-balance argument.
    The group of Ringeval and al.~is currently working to improve the measurement of the loop production function in their own simulations to see whether an agreement can be met and results of \cite{Blanco-Pillado:2019tbi} can be reproduced or not.

    For the remainder of this paper, for a given value for $\chi$, we will determine the normalization factor $c$ as to fit the parameter $A$ of the large scale \gls{lnd}.
    This assumption allows us to study both models on the same footing and is more likely to remain valid once an agreement will be found.

\subsection{Decomposition of the contributions in the different eras}
    \label{sec:decompose}

    \begin{figure}
        \centering
        \includegraphics{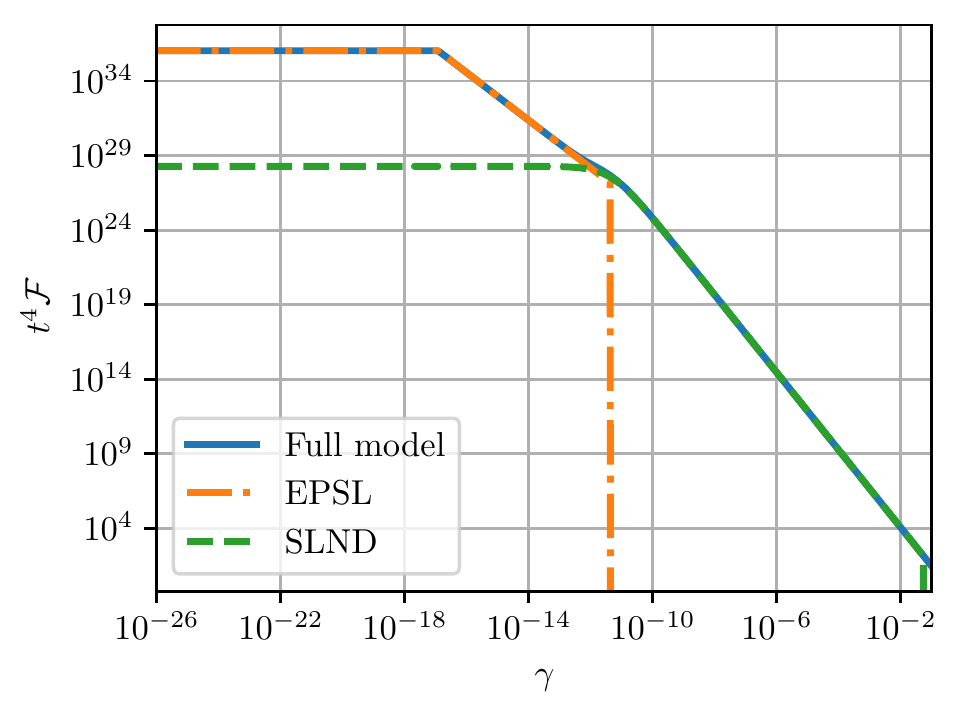}
        \caption{The decomposition of the \gls{lnd} into two populations, the \gls{slnd} and the \gls{epsl} for $\chi=0.2$ and $G\mu = 10^{-13}$ in the radiation era.
        The infrared cutoff is set to $\gammai = 0.1$.}
        \label{fig:decomposition}
    \end{figure}

        The aim of this study is to determine whether the \acrfull{epsl} described in \cite{lorenz_cosmic_2010,auclair_cosmic_2019} are observable features of the \gls{sbgw}.
        To this end, we propose a natural decomposition of the loop number density into two parts, as figure \ref{fig:decomposition} illustrates.
        The first contribution, which we called the \acrfull{slnd}, is of the form
        \begin{equation}
        t^4 \calF(\gamma) = C (\gamma+\gammad)^{-p} \Theta(\gammai - \gamma)
        \label{eq:lld}
        \end{equation}
        where $\gammai$ is a cutoff on the sizes of the loops.
        It is, for instance, the result of a Dirac loop production $t^5 \calP = c \delta(\gammai - \gamma)$ \cite{vilenkin_cosmic_2001}.
        In this particular case $C = c (\gammai+\gammad)^{3-3\nu}$ and $p=4-3\nu$.
        It also describes well the large scale behavior if the loop production function is the power-law of equation \eqref{eq:looprod} \cite{auclair_cosmic_2019}.
        Then the constants are fixed by
        \begin{itemize}
            \item in the sub-critical regime $\chi < \chicrit$, $C=\dfrac{c}{\meps}$ and $p=3-2\chi$
            \item in the super-critical regime $\chi > \chicrit$, $C=-\dfrac{c\gammai^{-\meps}}{\meps}$ and $p=4-3\nu$
        \end{itemize}
        where $\meps$ is given in equation \eqref{eq:meps} and $c$ is fixed by the normalization of the loop production function, as discussed in section \ref{sec:norm}.
        These approximations break down near $\chicrit$ and one should add regularization terms coming directly from the analytical expression of \cite{auclair_cosmic_2019}.
        For clarity we omit these terms here and put the details in appendix \ref{sec:regul}.

        On top of the \gls{slnd}, we superimpose an \acrfull{epsl} described as a piece-wise function, motivated by the work of \cite{auclair_cosmic_2019}
        \footnote{Based only on the asymptotic description provided in table 2 of \cite{auclair_cosmic_2019} this decomposition might seem artificial and one could concerned that loops smaller than $\gammad$ are counted twice.
        In fact it is just the opposite and this decomposition is well motivated when we refer to the full solutions.
        See appendix \ref{sec:note} for more details.}
        \begin{equation}
        t^4 \calF(\gamma) = \left\{
        \begin{array}{ll}
        \dfrac{c \gammad^{-1}}{2-2\chi} \gammac^{2\chi-2} & \text{ if } \gamma < \gammac\\
        \dfrac{c \gammad^{-1}}{2-2\chi} \gamma^{2\chi-2} & \text{ if } \gammac < \gamma <\gammad\\
        0  & \text{ if } \gammad < \gamma
        \end{array}
        \right. .
        \label{eq:sld}
        \end{equation}
        This definition comes directly from the fact that we assumed a sharp cutoff at the \emph{back-reaction scale} $\gammac$.
        The analytic formulae would be a little more complicated with a power-law cutoff, but the result would not be qualitatively modified.

        In the following, the analysis focuses on the impact of these two populations either in radiation-dominated era, or in matter-dominated era.
        One should note that large loops produced during the radiation era can survive long enough to be an important source of \gls{gw} in the matter era.
        They are a non-scaling population of loops and some models (see \cite{blanco-pillado_stochastic_2017}) predict they dominate during the matter-dominated era.
        Their contribution to the \gls{sbgw} is calculated in Appendix \ref{sec:decay} and taken into account in our analysis.
        On the contrary loops of size smaller than $\gammad$ during radiation era, which is the case of the \gls{epsl}, do not survive long enough in the matter era to be a significant contribution to the \gls{sbgw}.

\section{The Stochastic Background of Gravitational Waves}
\label{sec:sbgw}

    \subsection{Emission of gravitational waves}
        Cosmic string loops oscillate and emit \gls{gw}.
        The incoherent sum of their gravitational radiation forms a \gls{sbgw} which was first calculated in \cite{vachaspati_gravitational_1985}.
        The oscillation of the loops is not the only channel of gravitational radiation and burst-like events, from cusps, kinks and kink-kink collisions are also sources of gravitational radiation whose wave-forms were calculated in \cite{Damour:2000wa,damour_gravitational_2001,ringeval_stochastic_2017}.

        There exists two main methods to calculate the \gls{sbgw}.
        The first consists in introducing an effective decomposition into harmonics $\Pm, \mathrm{m} \in \mathbb{N}$ where the lowest modes are dominated by the oscillatory movement of the loop with typical frequency $2/\ell$, where $\ell$ is the invariant length of the loop, and the higher modes are dominated by burst-like events \cite{blanco-pillado_stochastic_2017}.
        Typically $\Pm \propto m^{-q}$ with $q=4/3$ (respectively $5/3,2$) for cusps (respectively kinks and kink-kink collisions).
        The energy density carried by the \gls{gw} per unit logarithmic interval of frequency is given by \cite{blanco-pillado_stochastic_2017}
        \begin{align}
        \rhoGW(t,f) &= G\mu^2 \sum_{m=1}^{\infty} C_m(f) \Pm \label{eq:omegam}\\
        C_m (f) &= \frac{2m}{f^2} \int_0^{z_*} \frac{dz}{\uH(z)(1+z)^6} \calF \left[\frac{2m}{(1+z)f},t(z)\right]
        \end{align}
        in which $\uH(z)$ is the Hubble parameter, $t(z)$ is the cosmic time, and $f$ is the frequency of the wave in the detector.
        Details on the cosmological parameters used in this paper are summarized in appendix \ref{sec:cosmo}.
        The redshift at which cosmic strings where formed is denoted by $z_*$, and
        it depends on the energy scale of the phase transition determined by the string tension.
        Considering the phase transition happened during the radiation era and that the temperature today is $T_0$, the redshift $z_*$ is given by
        \begin{equation}
            1+z_* \propto \dfrac{10^{39/2} \text{ GeV} }{T_0} \sqrt{G\mu}
        \end{equation}
        which we will fix to be infinity in the following.

        The other method to calculate the \gls{sbgw} consists in considering the sum of all burst-like events which are typically not isotropic \cite{damour_gravitational_2001,siemens_gravitational_2006,olmez_gravitational-wave_2010}.
        This approach allows one to remove events resolved inside a detector from the \gls{sbgw}, as they are not part of the background anymore.
        A detailed discussion of the differences of the two approaches can be found in \cite{Auclair:2019wcv}.

        In this paper we will use the first method.
        To keep the following analysis simple, we make the simplifying assumption that cosmic string loops emit only in their fundamental mode.
        The modes $m>1$ are only a small modification of its qualitative properties \cite{sanidas_constraints_2012, Sousa:2020sxs} and we discuss briefly their impact section \ref{sec:fundamental}.
        Introducing $\Q = 16\pi / (3\Gamma)$, the fraction of the critical density given by the energy of \gls{gw} is
        \begin{equation}
        \OmegaGW(\ln f) = \dfrac{\Q}{f \uH_0^2}\gammad^2 \int_0^\infty \dfrac{\ud z}{\uH(z)(1+z)^6} \calF\left[\dfrac{2}{(1+z)f},t(z)\right].
        \label{eq:omega}
        \end{equation}

    \subsection{Asymptotic description of the stochastic background of GW}
        \label{sec:asympt}
        With the assumptions made in this framework, one can calculate the energy density power spectrum  for each contribution individually, namely the contribution from \gls{slnd} on one side and the contribution from the \gls{epsl} on the other side.
        Consider for instance the \gls{sbgw} produced by the \gls{slnd} in the radiation era.

        In the radiation era, we can make the following approximations for the Hubble parameter and the cosmic time
        \begin{align}
            \uH(z) &= (1+z)^2 \uHr \\
            t(z) &= \dfrac{1}{2(1+z)^2\uHr}
        \end{align}
        where $\uHr = \uHo \sqrt{\OmegaR}$.
        This allows us to simplify equation \eqref{eq:omega} into
        \begin{equation}
            \OmegaGW(\ln f) = \dfrac{64 \Q \uHr \OmegaR}{f}\gammad^2 \int_\zeq^\infty\ud z t^4\calF\left[\dfrac{4 (1+z) \uHr}{f}\right].
        \end{equation}
        Inserting the \gls{slnd} contribution from equation \eqref{eq:lld} and noticing that $p>1$
        \begin{equation}
            \OmegaGW(\ln f) = \dfrac{4 \Q C \OmegaR}{ (p-1)}\gammad^{3-p} \left[ \left(1+\dfrac{4\uHr(1+\zeq)}{f\gammad} \right)^{1-p}-\left(1+\dfrac{\gammai}{\gammad} \right)^{1-p}\right]
            \label{eq:omegarad}
        \end{equation}
        We can make several remarks on this particular result that can be extended to the other contributions.
        The power spectrum has of two characteristic frequency scales.
        In particular $f =  4\uHr(1+\zeq)\gammai^{-1}$ is a low frequency cutoff for the energy density.
        This frequency is so low with respect to the frequency range of the \gls{gw} detectors that we omit it in the following.
        The frequency $f = 4\uHr(1+\zeq)\gammad^{-1}$ is a knee in the \gls{sbgw}.
        These two scales are well separated and the power spectrum can be approximated by power-laws far from these frequencies.

        We performed the same calculations for the other contributions, the \gls{slnd} and \gls{epsl} during the radiation and the matter era in the Appendices \ref{sec:calc-rad} and \ref{sec:calc-mat} and summed up the asymptotic behavior in tables \ref{table:sub}, \ref{table:super} and \ref{table:small}.
        We can make the general remarks:
        \begin{itemize}
            \item a typical frequency scale at which the power spectrum presents a knee, roughly $\uHo \gammad^{-1}$ for the \gls{slnd} and $\uHo \gammac^{-1}$ for the \gls{epsl}. Those two frequencies are very well separated.
            \item at low and high frequencies, the power spectrum behaves as a power law
            \item the width of the knees can be estimated from the complete calculations but is essentially small compared to the separation between $\uHo \gammad^{-1}$ and $\uHo \gammac^{-1}$ for $G\mu \ll 1$
            \item the power spectrum is cutoff at low frequencies, roughly $\uHo \gammai^{-1}$ for the \gls{slnd} and $\uHo \gammad^{-1}$ for the \gls{epsl}
        \end{itemize}
        \begin{table}[ht!]
        \centering
        \renewcommand{\arraystretch}{2}
        \begin{tabular}{|c|c|c|}
        \hline
        Frequency range & $f \ll \uHo \gammad^{-1}$ & $\uHo \gammad^{-1} \ll f $\\
        \hline
        Radiation era & $\Qr\gammad^2 \left[\dfrac{f}{4(1+\zeq)\uHr}\right]^{2-2\chir}$ &  $\Qr \gammad^{2\chir}$ \\
        Matter era & $\dfrac{\Qm}{ 2 (2-\chim)} \gammad^2 \left(\dfrac{f}{3\uHm} \right)^{2-2\chim}$& $3\uHm \Qm\gammad^{2\chim-1} f^{-1}$ \\
        Decaying into matter era & $\dfrac{2 \Qrm (1+\zeq)^{2\chir}}{(4\chir+1)} \gammad^2 \left(\dfrac{f}{3\sqrt{1+\zeq}\uHm}\right)^{2-2\chir} $ &$\dfrac{3 \Qrm \uHm}{(1-3\chir)} \gammad^{2\chir-1}f^{-1}$\\
        \hline
        \end{tabular}
        \caption{\gls{slnd} -- sub-critical case. For clarity, we have introduced $\Qr = \dfrac{4 \ucr \OmegaR \Q}{(1-\chir)(1-4\chir)}$, $\Qm = \dfrac{27 \ucm \OmegaM \Q}{8(1-2\chim)}$ and $\Qrm = \dfrac{27 \ucr \OmegaM \Q}{8(1-4\chir)\sqrt{1+\zeq}}$}
        \label{table:sub}
        \end{table}


        \begin{table}[ht!]
        \centering
        \renewcommand{\arraystretch}{2}
        \begin{tabular}{|c|c|c|}
        \hline
        Frequency range & $f \ll \uHo \gammad^{-1}$ & $\uHo \gammad^{-1} \ll f $ \\
        \hline
        Radiation era &$ \tQr \gammad^{2} \left(\dfrac{f}{4(1+\zeq)\uHr}\right)^{3/2}$ & $\tQr \gammad^{1/2}$ \\
        Matter era &$\dfrac{\tQm}{\uHm}\gammad^{2} f$ & $3 \uHm \tQm f^{-1}$\\
        Decaying into matter era & $\tQrm \gammad^{2} \left(\dfrac{f}{3\sqrt{1+\zeq}\uHm}\right)^{3/2} $ & $\dfrac{12 \uHm \tQrm}{\sqrt{1+\zeq}} \gammad^{-1/2} f^{-1}$\\
        \hline
        \end{tabular}
        \caption{\gls{slnd} -- super-critical case. For clarity, $\tQr = \dfrac{16 \ucr \gammai^{-\mur} \OmegaR \Q}{3(4\chir-1)}$, $\tQm = \dfrac{3 \ucm \OmegaM \gammai^{-\mum}\Q}{8(2\chim-1)}$ and $\tQrm = \dfrac{27 \ucr \OmegaM \gammai^{-\mum}\Q}{8(4\chir-1)}$}
        \label{table:super}
        \end{table}

        From these tables, one recovers that the loops produced during the radiation-dominated era give a plateau at high frequencies while all the other contributions decay as $f^{-1}$ meaning that at high enough frequencies, the \gls{sbgw} is a plateau where the dominant contribution comes from the radiation era.
        On the contrary, the low frequency region is usually dominated by \gls{gw} produced during the matter-dominated era.
        Indeed, the contributions from radiation era and from the loops produced in radiation era and decaying into matter era have similar shapes in the low frequency range, but as $\OmegaM \gg \OmegaR$, the latter contribution dominates.

        Another feature one can see is that in the sub-critical case (table \ref{table:sub}), the slopes of the \gls{sbgw} from the large loop population is dependent on the values of $\chir$ and $\chim$ where the $\ur$ index denotes radiation-domination and $\um$ matter-domination.
        Whereas in the super-critical regime table \ref{table:super} the frequency dependence of the spectrum is completely frozen.

        For the \gls{epsl}, the spectrum presents a knee at the frequency scale $\uHo \gammac^{-1}$ and is completely suppressed on frequencies below $\uHo \gammad^{-1}$.
        Therefore, any impact on the \gls{sbgw} happens on frequencies higher than $\uHo \gammad^{-1}$.
        In this frequency range, the dominant contribution coming for \gls{slnd} is the radiation-domination one.


        \begin{table}[ht!]
        \centering
        \renewcommand{\arraystretch}{2}
        \begin{tabular}{|c|c|c|}
        \hline
        Frequency range & $f \ll \uHo \gammac^{-1}$ & $\uHo \gammac^{-1} \ll f $ \\
        \hline
        Radiation era $\chir < 1/2$ & $\dfrac{\hQr}{(1-2\chir)(1-\chir)} \gammad \left[\dfrac{f}{4(1+\zeq)\uHr}\right]^{1-2\chir} $& $\dfrac{2 \hQr}{(1-2\chir)} \gammad^{-1}\gammac^{2\chir-1}$ \\
        Radiation era $\chir = 1/2$ &$2 \hQr \gammad \ln\left(\dfrac{\gammad f}{4\uHr(1+\zeq)}\right)$ &$2 \hQr\left[1+\ln\left(\dfrac{\gammad}{\gammac}\right)\right] \gammad$  \\
        Radiation era $\chir > 1/2$ & $\dfrac{\hQr}{(2\chir-1)(1-\chir)} \gammad^{2\chir}$ & $\dfrac{\hQr}{(2\chir-1)(1-\chir)} \gammad^{2\chir}$\\
        Matter era & $ \dfrac{\hQm}{3-2\chim} \gammad \left(\dfrac{f}{3\uHm}\right)^{1-2\chim}$ &$3\uHm \hQm \gammad \gammac^{2\chim-2} f^{-1}$\\
        \hline
        \end{tabular}
        \caption{\gls{epsl}. For clarity, $\hQr = 2\ucr \OmegaR \Q$ and $\hQm = \dfrac{27 \ucm \OmegaM Q}{16(1-\chim)}$}
        \label{table:small}
        \end{table}


    \subsection{Beyond the fundamental mode}
    \label{sec:fundamental}

        In subsection \ref{sec:asympt} we have made the assumption that a loop emits \gls{gw} in its fundamental mode, but this is not generally the case, especially if cusps or kinks are present on the loop \cite{damour_gravitational_2001,siemens_gravitational_2006}.
        If cusps or kinks are present, the higher modes of the spectral power $\Pm$ are not zero but behave as $m^{-q}$ where $q=4/3$ for cusps, $5/3$ for kinks and $2$ for kink-kink collisions.
        Even though there have been attempts to calculate the spectral power for all values of $m$ \cite{allen_gravitational_1992}, some even taking into account the gravitational back-reaction \cite{blanco-pillado_gravitational_2018}, we will make the following Ansatz for $\Pm$
        \begin{equation}
            \Pm = \Gamma \dfrac{m^{-q}}{\zeta(q)}
        \end{equation}
        where $\zeta$ is the Riemann zeta function to ensure the normalization of $\Pm$.
        Starting from equation \eqref{eq:omegam} during the radiation era and injecting this spectral power $\Pm$ gives
        \begin{equation}
            \OmegaGW(\ln f) = \dfrac{16\Q  \uHr \OmegaR}{f}\gammad^2 \sum_1^\infty \dfrac{m^{-q}}{\zeta(q)} m \int_\zeq^\infty\ud z t^4\calF\left[\dfrac{4 (1+z) m \uHr}{f}\right].
        \end{equation}
        For the the \gls{slnd} from equation \eqref{eq:lld}
        \begin{equation}
            \OmegaGW(\ln f) = \dfrac{4 \Q C \OmegaR}{ (p-1)}\gammad^2 \sum_1^\infty \dfrac{m^{-q}}{\zeta(q)} \left[ \left(\dfrac{4\uHr m (1+\zeq)}{f} +\gammad\right)^{1-p}-(\gammai + \gammad )^{1-p}\right].
        \end{equation}
        At high frequency and under the assumption that $\gammai \gg \gammad$, the spectral power is factorized and one recovers the result assuming only the fundamental mode
        \begin{equation}
            \OmegaGW(\ln f) = \dfrac{4\Q C \OmegaR}{(p-1)}\gammad^{3-p}.
        \end{equation}
        At low frequency the picture is slightly different and
        \begin{equation}
            \OmegaGW(\ln f) = \dfrac{4\Q C \OmegaR}{(p-1)}\gammad^2 \left[\dfrac{4\uHr (1+\zeq)}{f}\right]^{1-p} \dfrac{\zeta(p+q-1)}{\zeta(q)}.
        \end{equation}
        Even though we have only included the effects of the spectral power $\Pm$ on this single case, a simple calculation shows that this result can be generalized to the other types of loops distribution we discussed so far.
        At high frequencies, the \gls{sbgw} of \gls{gw} is insensitive to the decomposition into harmonics, while at low frequencies it is multiplied by a factor
        \begin{equation}
            \dfrac{\zeta(p+q-1)}{\zeta(q)}.
        \end{equation}

\section{Results}
\label{sec:results}

The aim of this section is to characterise the shape of the \gls{sbgw}, as a function of the loop production function exponents $\chir$ and $\chim$.
In particular, we assess the influence of the \gls{epsl} on the \gls{sbgw} and divide the parameter space $(\chir, \chim)$ into four classes with specific features.
    \subsection{Influence of the Extra Population of Small Loops on the SBGW}
        \begin{figure}
            \centering
            \includegraphics{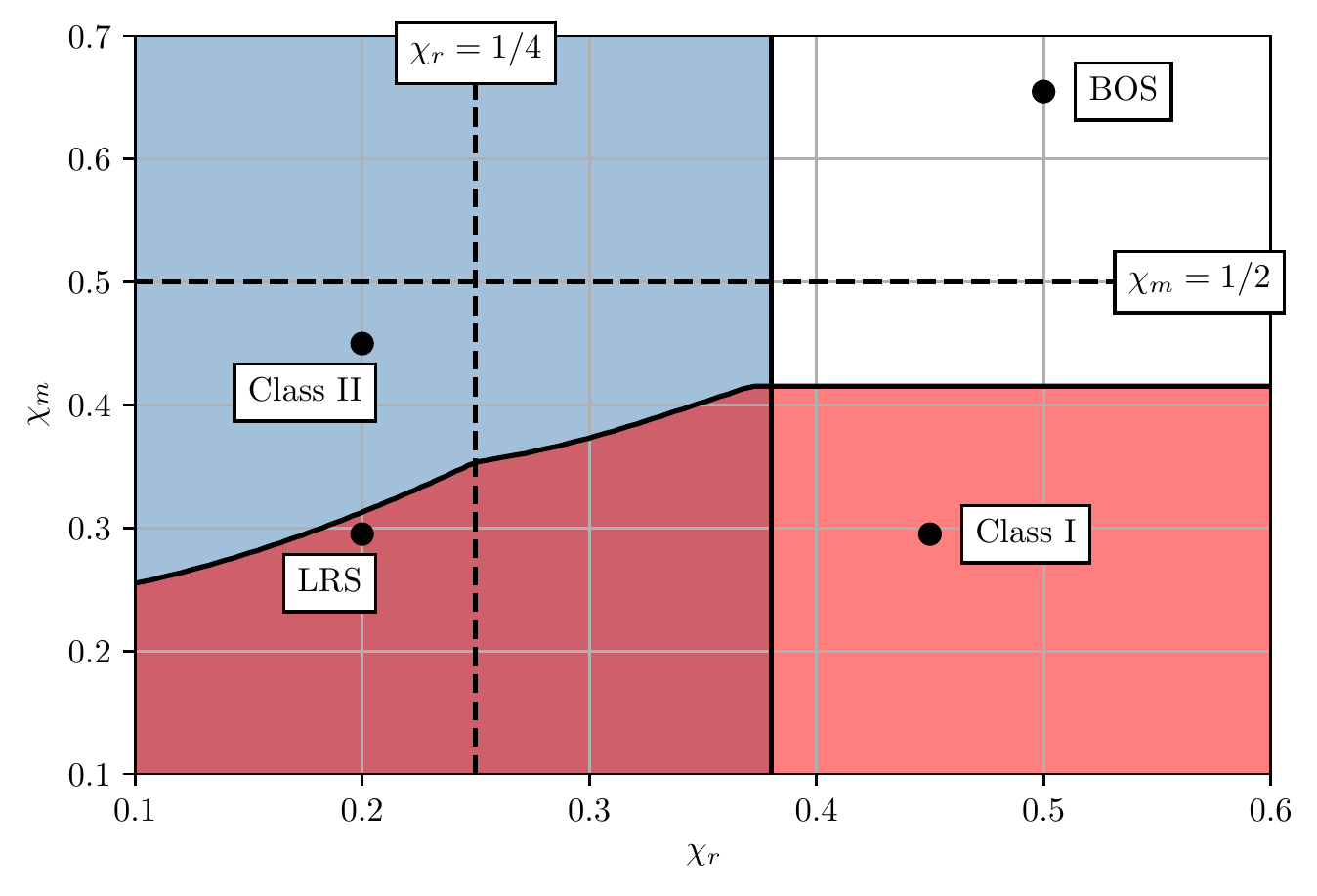}
            \caption{Impact of the extra population of small loops onto the \gls{sbgw} in the parameter space $\chir,\chim$ for $G\mu=10^{-13}$. In the blue region, the high frequency plateau for $\OmegaGW$ is dominated by the extra population of small loops produced during radiation era. In the red region, the spectrum presents a peak around $\uHo {\gammac^{(m)}}^{-1}$ produced by the \gls{epsl} during the matter era. Outside those regions, the population of small loops can be neglected.}
            \label{fig:regions}
        \end{figure}

        We can split the parameter space $(\chir,\chim)$ in different regions depending on whether the \gls{epsl} from radiation or matter era has a significant imprint on the \gls{sbgw}.

        Loops from the radiation era produce a plateau at high frequency in the \gls{sbgw}.
        The extra population of small loops introduces new features in the spectrum if its plateau is higher than the plateau of \gls{slnd}, meaning
        \begin{equation}
            \dfrac{3(2\chir-1/2)}{2(1-2\chir)} \left(\dfrac{\Upsilon}{\gammai}\right)^{2\chir-1}  \sqrt{\dfrac{\Gamma}{\gammai}}(G\mu)^{4\chir^2-1/2} > 1.
        \end{equation}
        This is shown as the blue region of figure \ref{fig:regions}.
        In this figure we have used the regularized formulae of appendix \ref{sec:regul} around $\chicrit$.
        We provide an analytical expansion in terms of $1/\ln(G\mu)$ in appendix \ref{sec:expansion-rad} for the position of the blue region.
        It should be noted that the \gls{epsl} produced during radiation era can be dominant at high frequencies even if the network is super-critical.
        This sets a new scale for $\chir$, between $\chicrit$ and $\chiir$.

        For loops produced during matter era, we assume that the extra population of small loops is visible if its peak at frequency $3\uHm\gammac^{-1}$ with amplitude
        \begin{equation}
            \dfrac{27\Q\ucm\OmegaM}{8(3-2\chim)(2-2\chim)}\gammad \gammac^{2\chim-1}
        \end{equation}
        is bigger than all the other contributions at this frequency.
        This is represented as the red region in figure \ref{fig:regions}.
        Contrary to the loops produced during the radiation era, only a subset of the sub-critical models during matter era produce detectable features for the \gls{sbgw}.

        From figure \ref{fig:regions}, one can see that the \gls{bos} model can be safely replaced by an effective Dirac distribution loop production function for two reasons.
        First, the network is super-critical during both matter and radiation era meaning the \gls{slnd} is universal with slope $-5/2$ during the radiation era and $-2$ during matter era \cite{auclair_cosmic_2019}.
        Secondly, figure \ref{fig:regions} shows that the extra population of small loops has a negligible impact on the \gls{sbgw}.

    \subsection{Hybrid models}

        \begin{figure}
            \centering
            \begin{subfigure}{0.49\textwidth}
                \caption{$G\mu = 10^{-13}$, $\chir = 0.5$, $\chim = 0.655$}
                \includegraphics[width=\textwidth]{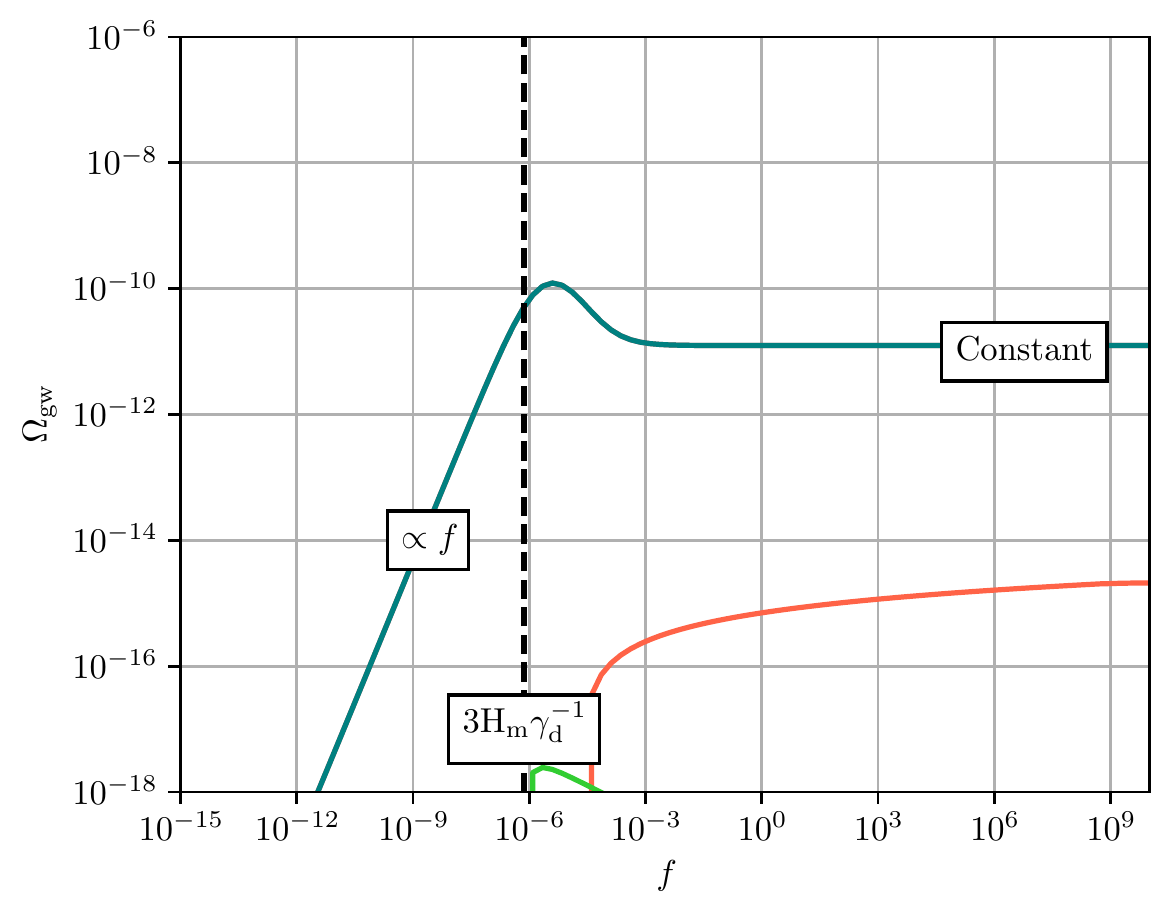}
                \label{fig:bos}
            \end{subfigure}
            \begin{subfigure}{0.49\textwidth}
                \caption{$G\mu = 10^{-13}$, $\chir = 0.2$, $\chim = 0.295$}
                \includegraphics[width=\textwidth]{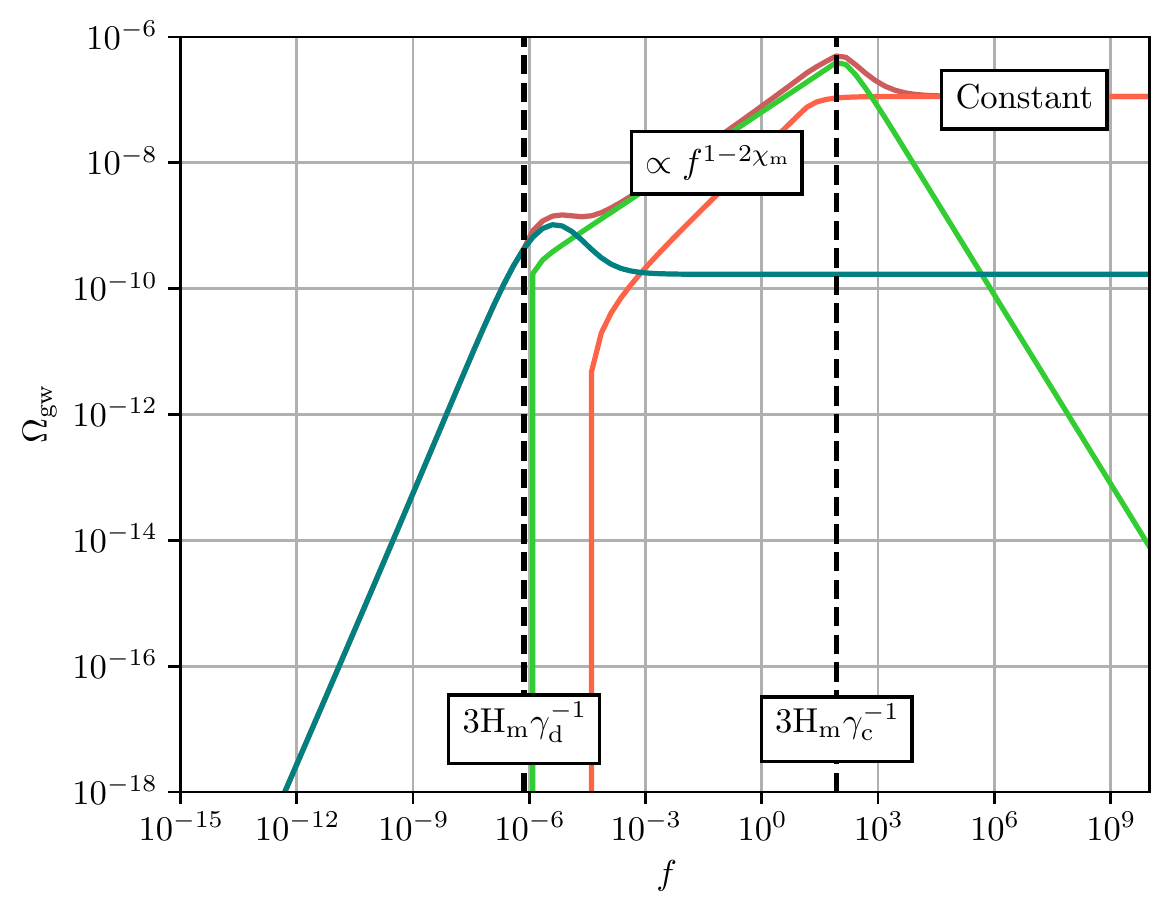}
                \label{fig:lrs}
            \end{subfigure}
            \begin{subfigure}{0.49\textwidth}
                \includegraphics[width=\textwidth]{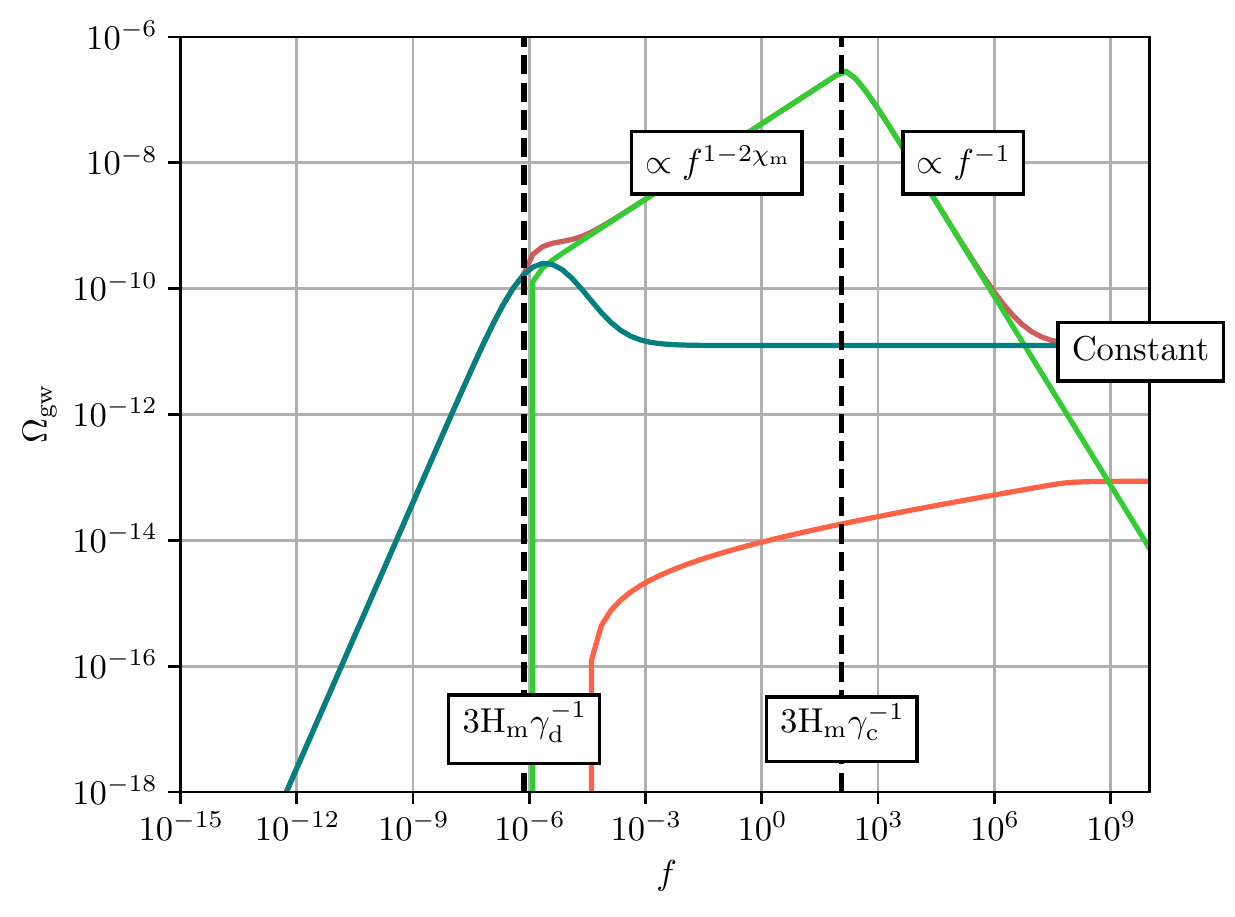}
                \caption{$G\mu = 10^{-13}$, $\chir = 0.45$, $\chim = 0.295$}
                \label{fig:classii}
            \end{subfigure}
            \begin{subfigure}{0.49\textwidth}
                \includegraphics[width=\textwidth]{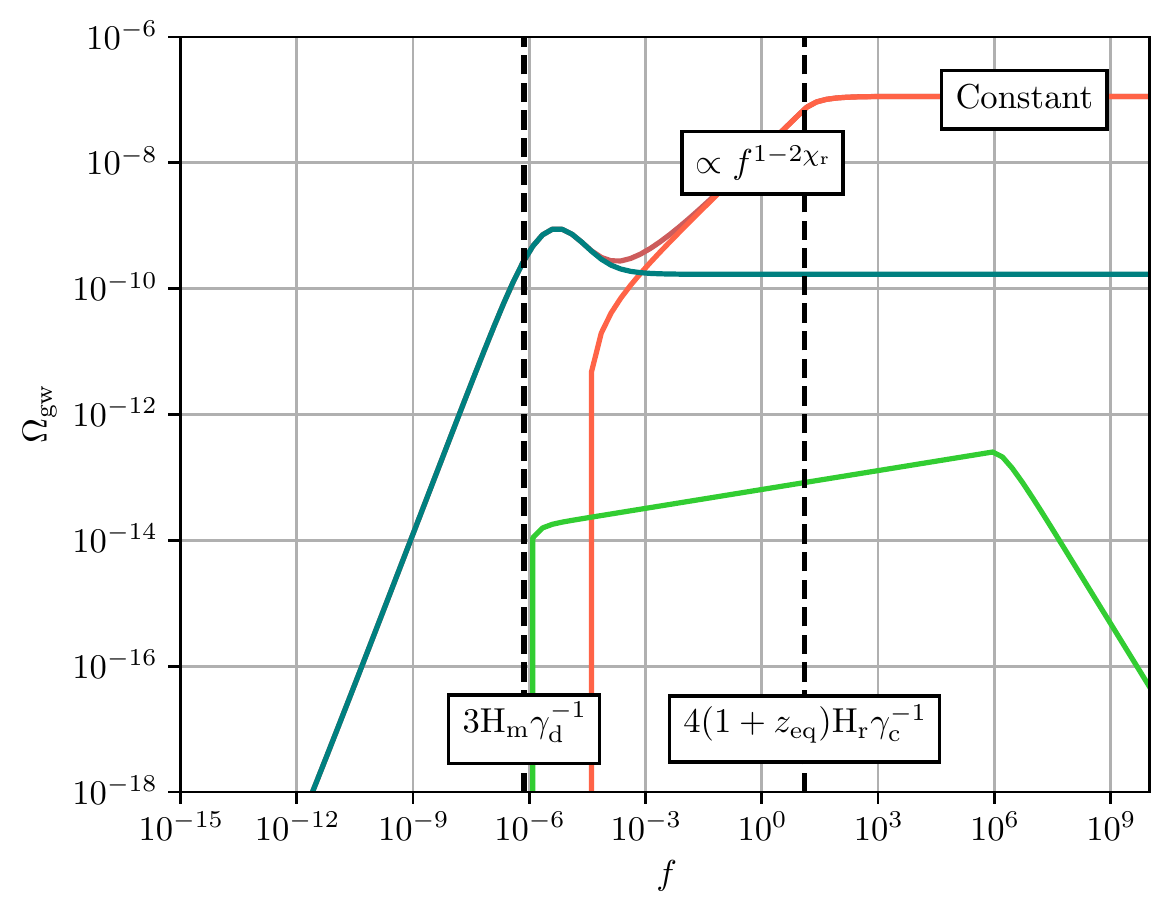}
                \caption{$G\mu = 10^{-13}$, $\chir = 0.2$, $\chim = 0.45$}
                \label{fig:classi}
            \end{subfigure}
            \caption{Four different classes of \gls{sbgw}}
            \label{fig:lnd}
        \end{figure}

        Figure \ref{fig:regions} can be used to build a classification of the various \gls{sbgw} in the parameter space $(\chir, \chim)$.
        Including the separation between sub-critical and super-critical regimes, there are nine different classes of spectra one can expect.
        For simplicity let us neglect the separation between sub-critical and super-critical and present four classes having distinctive features in terms of the \gls{sbgw}.

        The two first classes are represented by the well-known \gls{bos} model in figure \ref{fig:bos} and the \gls{lrs} model in figure \ref{fig:lrs} whose properties have been summed up on the figure.
        As we showed in the previous section, the \gls{bos} model can effectively neglect entirely the \gls{epsl}.
        On the contrary, it \gls{epsl} is a dominant source of \gls{gw} in both the radiation and the matter era for the \gls{lrs} model.

        We can add to this list two new hybrid classes of models.
        In figure \ref{fig:classii}, the \gls{epsl} of the radiation era can be neglected but not during the matter era, leading to peak around the frequency $3\uHm \gammac^{-1}$.
        As we explain in the following section, this peak leads to interesting features when we consider the detection by \gls{gw} detectors.
        Figure \ref{fig:classi} shows the opposite class in which the \gls{epsl} of the matter era can be neglected but not in the radiation era, producing a small valley in the \gls{sbgw}.

        As we attempted to make apparent in figure \ref{fig:lnd}, each of those classes have different shapes on which one can read the parameters of the cosmic string network, apart from models like the \gls{bos} models, for which the shape of the \gls{sbgw} does not depend on $(\chir,\chim)$.

    \subsection{Constraints on the string tension from GW experiments}
    \label{sec:gwdetect}

        We have not yet been able to detect any \gls{sbgw} in the European Pulsar Timing Array \cite{sanidas_constraints_2012} nor in the first two LIGO/Virgo runs\cite{abbott_constraints_2018,the_ligo_scientific_collaboration_search_2019}, giving only upper bounds on the cosmic string tension.
        New data analysis techniques are being devised for the next generation of \gls{gw} detectors such as LISA \cite{caprini_reconstructing_2019}.
        If ongoing and future \gls{gw} experiments could potentially detect the \gls{sbgw} coming from cosmic strings, it is a challenging data analysis problem to characterize the observed spectrum and distinguish between the variety of expected astrophysical and cosmological sources.

\begin{figure}
        \centering
        \begin{subfigure}{0.49\textwidth}
            \includegraphics[width=\textwidth]{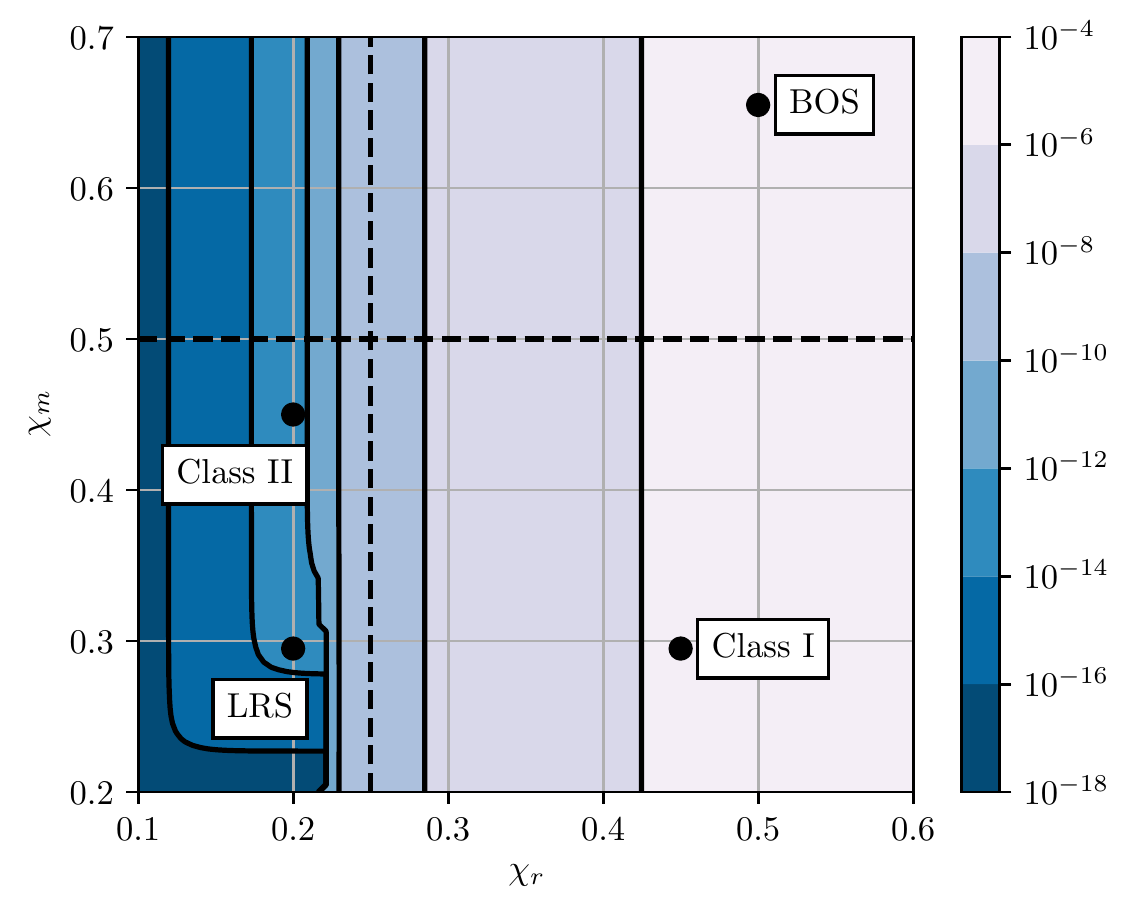}
            \caption{LIGO/Virgo, sensitivity of $\OmegaGW=10^{-7}$ taken at $f=20$ Hz}
            \label{fig:ligo}
        \end{subfigure}
        \begin{subfigure}{0.49\textwidth}
            \includegraphics[width=\textwidth]{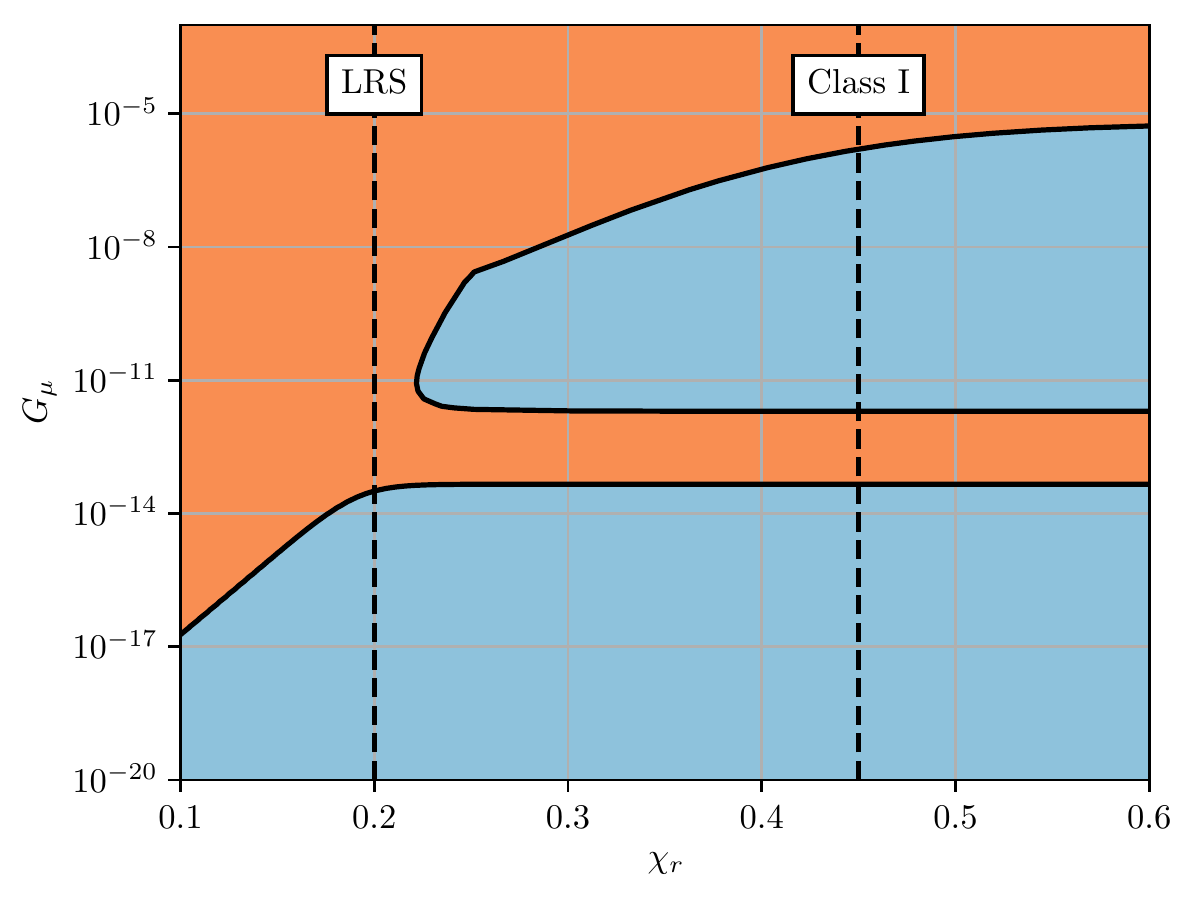}
            \caption{LIGO/Virgo constraints on $G\mu$ at $\chim=0.295$. The orange region is excluded giving non-convex constraints on $G\mu$ for a given model.}
            \label{fig:ligobranch}
        \end{subfigure}
        \begin{subfigure}{0.49\textwidth}
            \includegraphics[width=\textwidth]{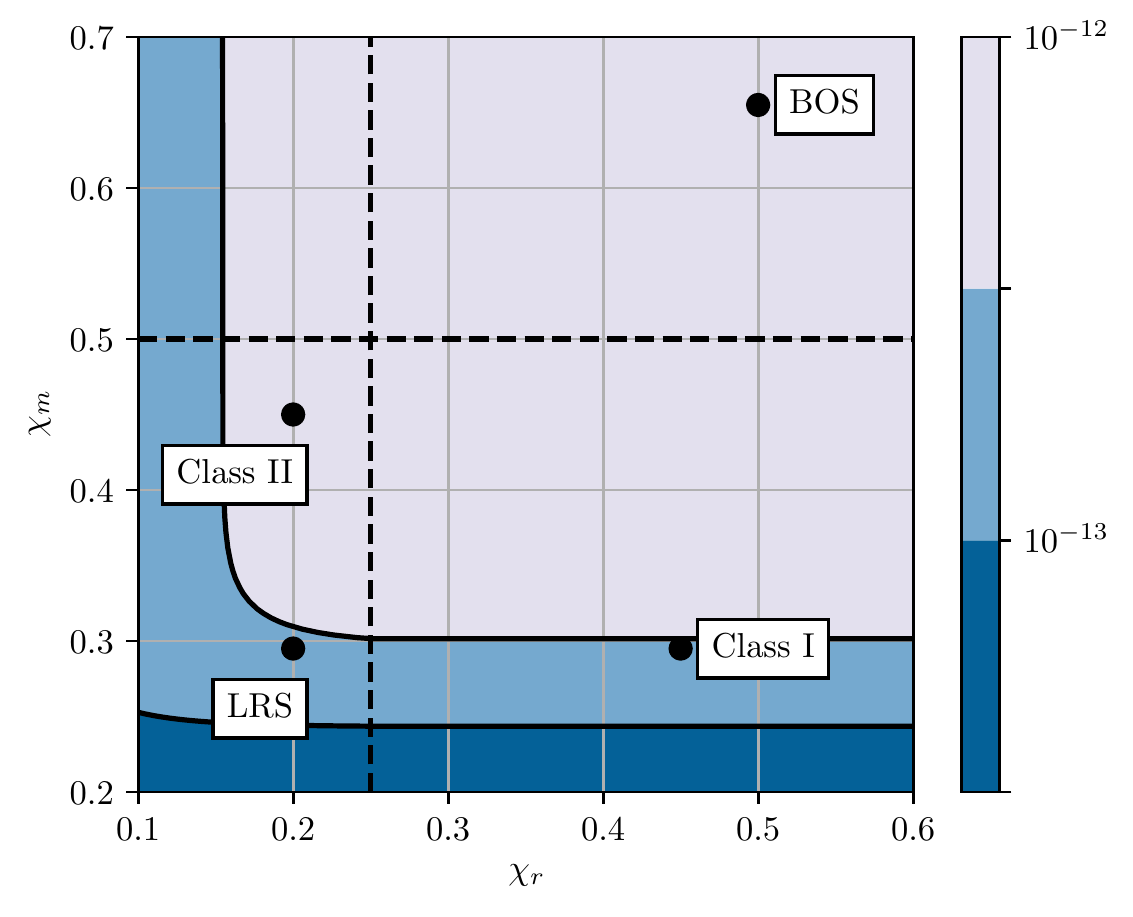}
            \caption{PTA, sensitivity of $\OmegaGW=10^{-12}$ taken at $f=2\times 10^{-9}$ Hz}
            \label{fig:pta}
        \end{subfigure}
        \begin{subfigure}{0.49\textwidth}
            \includegraphics[width=\textwidth]{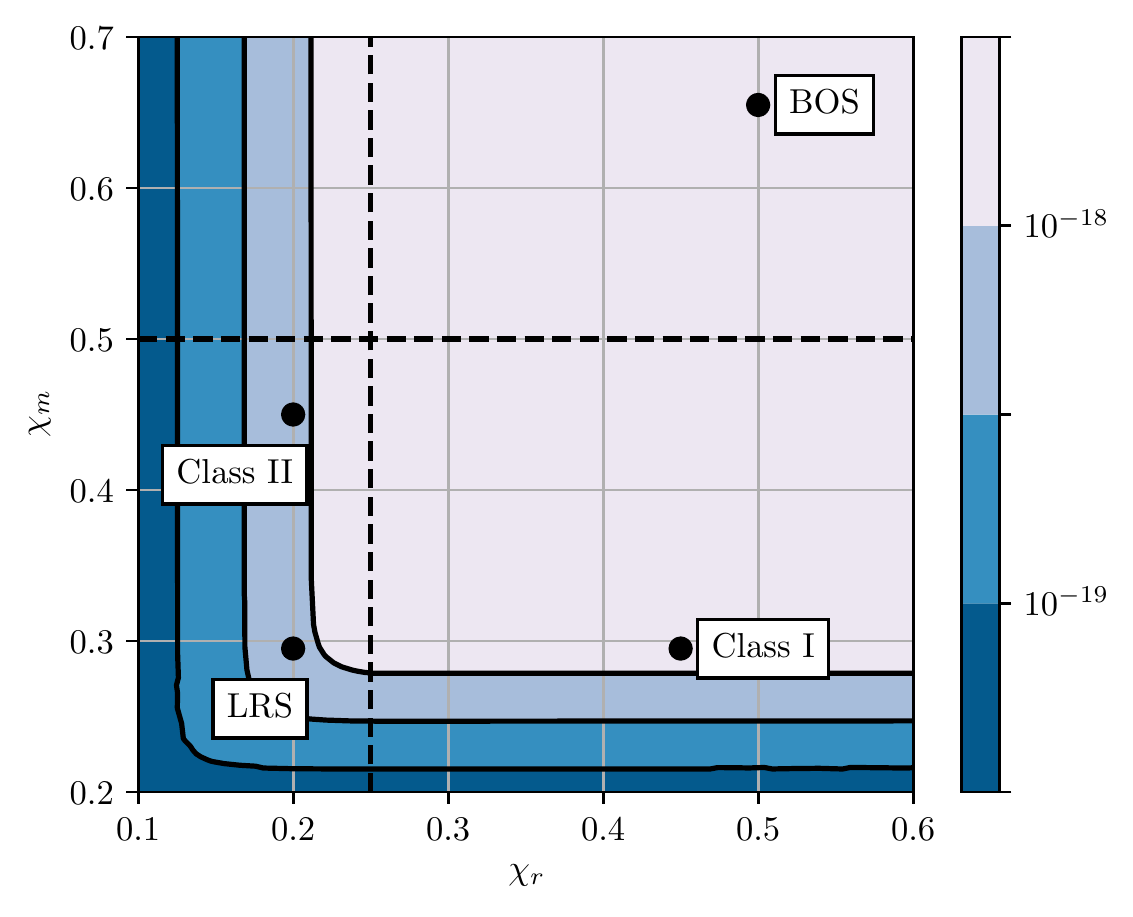}
            \caption{LISA, sensitivity of $\OmegaGW=10^{-13}$ taken at $f=10^{-2}$ Hz}
            \label{fig:lisa}
        \end{subfigure}
        \caption{Detection surface for the three types of \gls{gw} detectors in the $(\chir, \chim)$ parameter space. The color scale gives the upper bound on $G\mu$. Note that the detection surface is folded for LIGO/Virgo explaining why constraints on $G\mu$ jump several orders of magnitude in the lower left corner. Figure \ref{fig:ligobranch} is a slice at constant $\chim$.}
        \label{fig:gwdetectors}
    \end{figure}

        In this section, we do not pretend to tackle any of the technical difficulties of the detection of a \gls{sbgw}. In particular we will assume that we are able to separate the astrophysical foreground from the cosmological source of \gls{gw}.
        The theoretical \gls{gw} detector is modeled as having a given sensitivity curve, function of the frequency.
        We will make the assumption that the bandwidth of the detector is infinitely thin around a typical frequency and a given sensitivity $\OmegaGW$.
        This is of course a brutal assumption, however we expect that progress in the data analysis techniques can be effectively taken into account by changing the sensitivity of the instrument.
        As we possess analytic expressions for the stochastic background of gravitational waves within our framework, we can easily explore the parameter space $(G\mu,\chir,\chim)$.
        The result are summarized in figure \ref{fig:gwdetectors}.

        As was shown in previous sections, the extra population of small loops modifies the \gls{gw} spectrum at frequencies higher than $3\uHm \gammad^{-1}$, hence we expect it to have an impact on high frequency instruments such as LIGO/Virgo.
        It turns out the effect of the \gls{epsl} is quite dramatic for ground-based telescopes as illustrated in figure \ref{fig:ligo}.
        Not only does the constraint on $G\mu$ spans over nearly 10 orders of magnitude on the parameter space, it also present a folding for small values of $\chim \lesssim 0.3$ and $\chir \gtrsim 0.3$.
        The folding is illustrated by a slice at constant $\chim$ in figure \ref{fig:ligobranch}.
        This peculiar feature means that the constraint on $G\mu$ for these models is not an upper bound on $G\mu$ but rather that a set of intervals for $G\mu$ being excluded.
        This can be understood by looking at figure \ref{fig:classii}.
        The peak at $f=3\uHm \gammad^{-1}$ caused by the \gls{epsl} produced during matter era enters within the bandwidth of the detectors for a given set of $G\mu$ excluding another interval for $G\mu$.

        On the contrary, experiments at lower frequencies, are not affected by the extra population of small loops and are only sensitive to the slopes of the \gls{slnd}.
        As the shape of the \gls{slnd} is universal for super-critical models we expect the detection surface to be flat in the upper-right corner for low frequency experiments.
        For sub-critical networks however, the shape of the spectrum is modified and we expect the detection surface to be dependent on the values of $\chir$ and $\chim$ as can be seen in figures \ref{fig:pta} and \ref{fig:lisa}.

\section{Conclusion}
\label{sec:conclusion}
    Our framework allowed us to produce analytic formulae for the \gls{sbgw} for cosmic strings including its small-scale structure.
    In particular, the introduction of a back-reaction scale $\gammac \ll \gammad$ produces an \acrfull{epsl} which can have an important effect on the \gls{sbgw} for the \gls{lrs} model \cite{lorenz_cosmic_2010}.
    We proposed a parametrization, using variables $\chir$ and $\chim$, of the uncertainty on the dynamics of the infinite string network \cite{lorenz_cosmic_2010,auclair_cosmic_2019}.
    We showed that the predictions of \gls{bos} \cite{blanco-pillado_stochastic_2017} are stable if one introduces this back-reaction scale, and that the extra population of loops is subdominant in terms of \gls{gw} production in this particular model.
    We are also in agreement with \gls{lrs} \cite{lorenz_cosmic_2010}.

    We showed the small-scale structure of cosmic strings can have a significant impact on the \gls{sbgw} even \emph{outside the super-critical regime} and calculated the region of the parameter space where its effect cannot be neglected.
    We classified the \gls{gw} power spectra coming from cosmic strings into four different classes, for which we have shown two new and called \emph{hybrid models}.
    The values of the parameters $\chir$ and $\chim$ for these two hybrid models are not supported by any numerical simulation, however the uncertainty on $\chir$ and $\chim$ motivates us to consider them.

    We have also estimated systematically the constraints on the string tension $G\mu$ of different types of \gls{gw} detectors and showed that low-frequency experiments will provide more stable and model-independent bounds while ground-based detectors will be very sensitive to the details of the small-scale structure of the cosmic string network.

\acknowledgments I would like to thank Dani\`ele Steer for her support and for useful discussions, and Christophe Ringeval, Mairi Sakellariadou and Jose Juan Blanco-Pillado for comments and questions on the first draft of this paper. I thank Nordita for hospitality whilst this work was in progress.

\bibliographystyle{JHEP}

\bibliography{strings}
\printglossary

\newpage
\appendix
\section{Cosmological parameters}
\label{sec:cosmo}
    We assumed a the $\Lambda$CDM cosmology with the parameters in table \ref{table:cosmo}.
    For the sake of simplicity, we neglected the impact of the late-time acceleration of the Universe.
    We also neglected the changes in the relativistic degrees of freedom, something which would decrease slightly the high frequency plateau.
    The Hubble parameter is
    \begin{equation}
        \uH(z) = \uHo \sqrt{\OmegaR (1+z)^4 + \OmegaM (1+z)^3}
    \end{equation}
    and the cosmic time is given by
    \begin{equation}
        t(z) = \int_z^\infty \dfrac{\ud z'}{\uH(z')(1+z')}
    \end{equation}

    \begin{table}[h!]
        \centering
        \begin{tabular}{|c|c|}
            \hline
            Parameter & Value \\
            \hline
            $h$ & $0.678$ \\
            $\uHo$ & $100 h \mathrm{km.s}^{-1} \mathrm{Mpc}^{-1}$ \\
            $\OmegaR$ & $9.1476 \times 10^{-5}$\\
            $\OmegaM$ & $0.308$\\
            $\uHr$ & $\uHo \sqrt{\OmegaR}$\\
            $\uHm$ & $\uHo \sqrt{\OmegaM}$\\
            \hline
        \end{tabular}
        \caption{Cosmological parameters from \cite{planck_collaboration_planck_2016}}
        \label{table:cosmo}
    \end{table}

\section{Note on the decomposition of the loop number density}
\label{sec:note}

In the sub-critical regime, the scaling \gls{lnd} is given by
\begin{equation}
    t^4 \calF(\gammac \leq \gamma) = \dfrac{c}{\meps} (\gamma + \gammad)^{2\chi-3}f\left(\dfrac{\gammad}{\gamma+\gammad}\right)
\end{equation}
In this equation the function $f$ is defined by
\begin{equation}
    f(x) \equiv \hypergauss{3-2\chi}{\meps}{\meps+1}{x} \underset{1}{\sim} \dfrac{\Gamma(3 \nu - 2\chi) \Gamma(2\chi-2)}{\Gamma(3\nu-3)} x^{-\meps} + \dfrac{\meps}{2-2\chi}
    \left(1-x\right)^{2\chi-2}
\end{equation}
where we have expanded the hypergeometric function around unity using Gamma functions \cite{gradshtein_table_2007}.
Taking the limit $\gammac \leq \gamma \ll \gammad$
\begin{equation}
    t^4 \calF(\gammac \leq \gamma \ll \gammad) \sim \dfrac{\Gamma(3 \nu - 2\chi) \Gamma(2\chi-2)}{\Gamma(3\nu-3)} \dfrac{c}{\meps} \gammad^{2\chi-3} + \dfrac{c\gammad^{-1}}{2-2\chi}
    \gamma^{2\chi-2}
\end{equation}
where the Gamma function factor is $1$ for $\chi=\chicrit$, of order unity for $0.1< \chi$ and eventually diverges for $\chi=0$.
A similar approach in the super-critical regime $\meps < 0$ leads to equation 3.12 of \cite{auclair_cosmic_2019}
\begin{equation}
t^4 \calF(\gammac \leq \gamma \ll \gammad) \simeq
 \dfrac{c\gammad^{-1}}{2-2\chi} \gamma^{2\chi-2} -\dfrac{c\gammai^{-\meps}}{\meps} \gammad^{-4+3\nu}
\end{equation}
The scale $\gammair$ giving a knee in the \gls{lnd} is precisely set by the competition between these two contributions.
It should be noted that the \gls{epsl} \emph{can be described uniformly in the three regimes} sub-critical, critical and super-critical.
This property makes it easier for us to conduct our analysis and makes this decomposition very natural.

\section{Regularization around $\chicrit$ for the standard loop number density}
\label{sec:regul}

    The decomposition for the \gls{slnd} of section \ref{sec:decompose} fails around $\chicrit$ and needs regularization terms to remain consistent.
    Introducing $\gammad$, we suggest the following scheme
    \begin{itemize}
        \item for sub-critical regimes, $\C = \dfrac{\uc}{\meps}\left[1-\left(\dfrac{\gammad}{\gammai}\right)^{\meps}\right]$
        \item for super-critical regimes, $\C = -\dfrac{\uc}{\meps}\left(\gammai^{-\meps}-\gammad^{-\meps}\right)$
    \end{itemize}
    leading to the limit when $\meps\rightarrow 0$
    \begin{equation}
        \C = \uc \ln\left(\dfrac{\gammad}{\gammai}\right)
    \end{equation}
    This regularization scheme gives a good approximation around $\chicrit$ at the expanse of underestimating a the \gls{lnd} for large $\gamma$.
    We can use this approximation to calculate the stochastic background for which we give the asymptotic behavior in table \ref{table:critical}.
    \begin{table}[ht!]
    \centering
    \renewcommand{\arraystretch}{2}
    \begin{tabular}{|c|c|c|}
    \hline
    Frequency range & $f \ll \uHo \gammad^{-1}$ & $\uHo \gammad^{-1} \ll f $ \\
    \hline
    Radiation era &$\Qr \gammad^{2} \left(\dfrac{f}{4(1+\zeq)\uHr}\right)^{3/2}$ & $\Qr \sqrt{\gammad}$ \\
    Matter era &$\dfrac{\Qm}{\uHm} \gammad^{2} f$ & $27 \uHm\Qm f^{-1}$ \\
    Decaying into matter era & $\dfrac{\Qrm}{4} \gammad^{2} \left(\dfrac{f}{3\sqrt{1+\zeq}\uHm}\right)^{3/2} $ & $\dfrac{3 \uHm\Qrm}{\sqrt{1+\zeq}} \gammad^{-1/2} f^{-1}$ \\
    \hline
    \end{tabular}
    \caption{\gls{slnd} -- critical case. For simplicity $\Qr = \dfrac{8 \ucr \OmegaR}{3} \ln\left(\frac{\gammai}{\gammad}\right)$, $\Qm = \dfrac{3 \ucm \OmegaM}{8}\ln\left(\frac{\gammai}{\gammad}\right)$, $\Qrm = \dfrac{27\ucr\OmegaM}{4}\ln\left(\frac{\gammai}{\gammad}\right)$}
    \label{table:critical}
    \end{table}

\section{Contributions in the radiation era}
\label{sec:calc-rad}
    In the radiation era, we can make the following approximations :
    \begin{align}
        \uH(z) &= (1+z)^2 \uHr \\
        t(z) &= \dfrac{1}{2(1+z)^2\uHr}
    \end{align}
    where $\uHr = \uHo \sqrt{\OmegaR}$.
    In this case,
    \begin{equation}
        \OmegaGW(\ln f) = \dfrac{64 \Q \uHr \OmegaR}{f}\gammad^2 \int_\zeq^\infty\ud z t^4\calF\left[\dfrac{4 (1+z) \uHr}{f}\right].
    \end{equation}

    \subsection{Standard loop distribution}

        If we consider in the radiation era a loop distribution function
        \begin{equation}
            t^4 \calF(\gamma)= C (\gamma+\gammad)^{-p} \Theta(\gammai-\gamma)
        \end{equation}
        we can evaluate analytically $\OmegaGW$.
        In particular there is a typical frequency $\fb = 4(1+\zeq)\uHr \gammad^{-1}$ which corresponds to a knee in the power spectrum which can be used to rewrite the \gls{gw} power spectrum.
        \begin{equation}
            \OmegaGW(\ln f) = \dfrac{4 \Q C \OmegaR}{ (p-1)}\gammad^{3-p} \left[ \left(1+\dfrac{4\uHr(1+\zeq)}{f\gammad} \right)^{1-p}-\left(1+\dfrac{\gammai}{\gammad} \right)^{1-p}\right]
        \end{equation}

    \subsection{Extra population of small loops}

        We perform the same analysis but with the distribution defined in equation \eqref{eq:sld}.
        Due to the piece-wise nature of the \gls{lnd}, we have to distinguish two cases.
        In this case, $\fa = 4(1+\zeq)\uHr \gammad^{-1}$ and $\fb = 4(1+\zeq)\uHr \gammac^{-1}$
        \begin{align}
            \OmegaGW(\ln f < \ln \fb) &= \dfrac{4 \Q \ucr \OmegaR}{(1-2\chir)(2-2\chir)}\gammad \gammac^{2\chir-1} \left[ \left(\dfrac{\fb}{f}\right)^{2\chir-1}-\left(\dfrac{\gammad}{\gammac}\right)^{2\chir-1} \right] \\
            \OmegaGW(\ln f > \ln \fb) &= \dfrac{4\Q \ucr \OmegaR}{(1-2\chir)(2-2\chir)}\gammad \gammac^{2\chir-1} \left[(2-2\chir) - \dfrac{\fb}{f}(1-2\chir) -\left(\dfrac{\gammad}{\gammac}\right)^{2\chir-1} \right]
            \label{eq:boost}
        \end{align}
        One can remark several things
        \begin{itemize}
            \item when $\chir < 1/2 $ the value of the plateau at high frequencies is given by the scale $\gammac$
            \item when $\chir > 1/2$ the plateau is given by the scale $\gammad$ in a way very similar to the \gls{slnd}
        \end{itemize}
        In the special case where $\chir=1/2$, the cutoff is of primordial importance
        \begin{align}
            \OmegaGW(\ln f < \ln \fb) &= 4\Q \ucr \OmegaR\gammad \ln\left(\dfrac{f}{\fa}\right)  \\
            \OmegaGW(\ln f > \ln \fb) &= 4\Q \ucr \OmegaR\gammad\left[1 - \dfrac{\fb}{f} +\ln\left(\dfrac{\gammad}{\gammac}\right)\right]
        \end{align}

\section{Contributions during matter era}
\label{sec:calc-mat}
    In the matter era, we can make the following approximations :
    \begin{align}
        \uH(z) &= (1+z)^{3/2} \uHm \\
        t(z) &= \dfrac{2}{3(1+z)^{3/2}\uHm}
    \end{align}

    where $\uHm = \uHo \sqrt{\OmegaM}$. In this case,
    \begin{equation}
        \OmegaGW(f) = \dfrac{81\Q\uHm \OmegaM}{16f} \gammad^2 \int_0^\zeq \ud z (1+z)^{-3/2} t^4\calF\left[\dfrac{3\sqrt{1+z}\uHm}{f}\right]
    \end{equation}

    We expect two types of sources in the matter era, scaling loops formed during the matter era and remnants from the radiation era which decay with time.

    \subsection{Scaling loops during matter era -- Standard loop distribution}

        Assuming a scaling large-loop distribution
        \begin{equation}
            t^4\calF(\gamma)= C (\gamma+\gammad)^{-p} \Theta(\gammai - \gamma)
        \end{equation}
        Changing variables from $z$ to $x=\dfrac{3\sqrt{1+z}\uHm}{{f\gammad}}$ we obtain
        \begin{equation}
            \OmegaGW(f) = \dfrac{3^5 \Q C\uHm^2 \OmegaM}{8 f^2}\gammad^{1-p} \int_{\frac{3\uHm}{{f\gammad}}}^{\frac{3\sqrt{1+\zeq}\uHm}{{f\gammad}}} \ud x  (1+x)^{-p} x^{-2} \Theta\left(\dfrac{\gammai}{\gammad}-x\right)
        \end{equation}

        \subsubsection{Approximate solution}

            One can introduce the typical frequency
            \begin{equation}
                \fc = (p+1)^{1/p} \dfrac{3\uHm}{\gammad}
            \end{equation}
            and use it to interpolate the \gls{gw} power spectrum between the two solvable regimes of low and high frequency
            \begin{equation}
                \OmegaGW(\ln f) = \dfrac{81\Q C \uHm \OmegaM}{8f}  \gammad^{2-p} \left(\dfrac{f}{\fc+f}\right)^p
            \end{equation}

        \subsubsection{Exact solution}

            There exists a well-defined exact primitive to this integral we can use to obtain an exact solution even around the peak.
            \begin{equation}
                -\dfrac{\hypergauss{p}{ 1 + p}{ 2 + p}{ -\dfrac{1}{x}}}{(1 + p) x^{p+1}}
            \end{equation}
            Indeed, even though the Gauss hypergeometric function has a radius of convergence of $1$, it turns out it converges for $ \dfrac{1}{x} < 0$.

            This primitive can be used but is not very practical.
            However we can use it to perform a simple comparison.
            We know that the region where the approximation will be the worse is around $\fc$, we can calculate the precision of this approximation there.

            In the case of the approximate solution
            \begin{equation}
                \OmegaGW(\fc) = \dfrac{3^4 \Q C  \uHm \OmegaM}{\fc 2^{p+3}}\gammad^{2-p}
            \end{equation}
            While for the exact solution, if $f = \fc$
            \begin{equation}
                \OmegaGW(\fc) = \dfrac{3^4\Q C \uHm \OmegaM}{8 \fc} \gammad^{2-p} \left[\hypergauss{p}{ 1 + p}{ 2 + p}{ -(p+1)^{1/p}}-\dfrac{\hypergauss{p}{ 1 + p}{ 2 + p}{ -\dfrac{(p+1)^{1/p}}{\sqrt{1+\zeq}}}}{\sqrt{1+\zeq}^{p+1}}\right]
            \end{equation}
            Then the ratio between  the exact value divided by the approximate one in the limit $\zeq \rightarrow \infty$ is
            \begin{equation}
                2^p \hypergauss{p}{ 1 + p}{ 2 + p}{ -(p+1)^{1/p}} \underset{p=2}{=} 0.82
            \end{equation}
            We see that the approximate solution overestimate the value of the peak.

    \subsection{Scaling loops during matter era -- extra population of small loops}

        Assuming the \gls{epsl} distribution. Five cases happen depending on the frequencies :
        \begin{align}
            \fone &= \dfrac{3\uHm}{\gammad}, \ftwo = \dfrac{3\uHm\sqrt{1+\zeq}}{\gammad}\\
            \fthre &= \dfrac{3\uHm}{\gammac}, \ffour = \dfrac{3\uHm\sqrt{1+\zeq}}{\gammac}
        \end{align}
        $\OmegaGW(f < \fone) = 0$. The other results come straightforwardly
        \begin{equation}
            \OmegaGW(\fone < f <\ftwo) = \dfrac{81 \Q\ucm \uHm\OmegaM }{8(3-2\chim)(2-2\chim)f}\gammad \gammac^{2\chim-2} \left(\dfrac{\fthre}{f}\right)^{2\chim-2} \left[1 - \left(\dfrac{\fone}{f}\right)^{3-2\chim} \right]
        \end{equation}
        we can check the continuity in $\fone$. The next region gives
        \begin{equation}
            \OmegaGW(\ftwo < f < \fthre) = \dfrac{81 \Q\ucm \uHm\OmegaM}{8(3-2\chim)(2-2\chim)f}\gammad \gammac^{2\chim-2} \left(\dfrac{\fthre}{f}\right)^{2\chim-2}\left[1 - \left(\dfrac{\fthre}{\ffour}\right)^{3-2\chim}\right]
        \end{equation}
        One can check easily the continuity in $\ftwo$ and $\fthre$.
        \begin{align}
            \OmegaGW(\fthre<f<\ffour) &=\dfrac{81 \Q\ucm \uHm \OmegaM}{8(3-2\chim)(2-2\chim)f}\gammad \gammac^{2\chim-2} \dfrac{\fthre}{f}\left[ (3-2\chim)\dfrac{f}{\fthre} + (2\chim-2) - \left(\dfrac{f}{\ffour}\right)^{3-2\chim} \right] \\
            &= \dfrac{81 \Q\ucm \uHm \OmegaM}{8(2-2\chim)f}\gammad\gammac^{2\chim-2}\left\{ 1 -\dfrac{\fthre}{f}\dfrac{1}{3-2\chim} \left[(2-2\chim) + \left(\dfrac{f}{\ffour}\right)^{3-2\chim}\right] \right\}
        \end{align}
        here are again two formulae for the continuity in $\fthre$ and $\ffour$.
        The last region gives
        \begin{equation}
            \OmegaGW(\ffour <f) = \dfrac{81 \Q\ucm \uHm \OmegaM}{8(2-2\chim)f}\gammad\gammac^{2\chim-2} \left(1 -\dfrac{\fthre}{\ffour}\right)
        \end{equation}

    \subsection{Decaying loops from radiation era}
    \label{sec:decay}

        For loops created during radiation era, the relaxation term in matter era is
        \begin{align}
            t^4\calF(\gamma)&= C \left(\dfrac{t}{\teq}\right)^4 \left(\dfrac{1+z}{1+\zeq} \right)^3 \left(\dfrac{\teq}{t}\right)^p (\gamma+\gammad)^{-p} \Theta\left[\gammai+\gammad -(\gamma+\gammad)\dfrac{t}{\teq}\right] \\
            &=c \left(\dfrac{1+\zeq}{1+z}\right)^{3-3p/2} (\gamma+\gammad)^{-p} \Theta\left[\gammai+\gammad -(\gamma+\gammad)\left(\dfrac{1+\zeq}{1+z}\right)^{3/2}\right]
        \end{align}
        Loops smaller than $\gammad$ decay very rapidly.
        Changing variables from $z$ to $x=\dfrac{3\sqrt{1+z}\uHm}{{f\gammad}}$
        \begin{equation}
            \OmegaGW(f) = \dfrac{81 \Q C \uHm \OmegaM}{8 f} (1+\zeq)^{3/2(2-p)} \gammad^{2-p} \left(\dfrac{f\gammad}{3\uHm}\right)^{3p-7} \int  x^{3p-8} (1+x)^{-p} \ud x
        \end{equation}

        \subsubsection{Approximate solution}

            Using the same idea as in the previous section, we introduce a new frequency $\fd$ that separates the different regimes.
            \begin{equation}
                \fd = \dfrac{3\uHm}{\gammad} \left[\dfrac{2p-7}{3p-7} \dfrac{(1+\zeq)^{-1/2} - (1+\zeq)^{3/2(2-p)}}{(1+\zeq)^{-(p+1)/2}- (1+\zeq)^{3/2(2-p)}} \right]^{1/p}
            \end{equation}
            One can use to find an interpolating formula for the \gls{gw} power spectrum
            \begin{equation}
                \OmegaGW(\ln f) = \dfrac{81 \Q C \uHm \OmegaM}{8f} \gammad^{2-p} \left(\dfrac{f}{f+\fd}\right)^{p}  \dfrac{(1+\zeq)^{-1/2} - (1+\zeq)^{3/2(2-p)}}{3p-7}
            \end{equation}

            This expression starts to be much more complicated because we need to keep track of the two boundary terms of the integral. Different behavior appear :
            \begin{itemize}
                \item $p < 7/3 \approx 2.33$, a very soft slope, all the integrals are dominated by the lower bound
                \item $7/3 < p < 7/2 = 3.5 $, the large $f$ bound is dominated by the lower bound and the low $f$ is dominated by the higher bound.
                \item $7/2 < p$, a very steep slope, all the integrals are dominated by the higher bound
            \end{itemize}
            In practice, we will only consider $p \in [2,3]$

            There is also a well defined primitive for this integral
            \begin{equation}
                \dfrac{x^{-7 + 3 p} \hypergauss{p}{ -7 + 3 p}{ -6 + 3 p}{ -x}}{-7 + 3 p}
            \end{equation}
            but it is not very practical to use.

\section{Analytic estimation for the boundary in $\chir$}
    \label{sec:expansion-rad}
        The question is to find for which values of $\chir$ does the \gls{epsl} leaves a signature in the \gls{sbgw}.
        This boils down to finding the value for $\chir$ at which the two contributions are equal.
        $G\mu$ being an infinitesimal quantity, one can perform an expansion as :
        \begin{equation}\chir(G\mu) = \chis + \dfrac{A}{\ln(G\mu)} + \dfrac{B}{\ln^2(G\mu)}\end{equation}.
        Where $\chis=\sqrt{3\nu-1}/2=\dfrac{1}{2\sqrt{2}}$. One obtains
        \begin{align}
        A &= \dfrac{1}{8\chis}\ln\left(\dfrac{(1-2\chis)}{(2\chis+1-3\nu)(3-3\nu)}\left(\dfrac{\Gamma}{\gammai}\right)^{3\nu-2}\left(\dfrac{\Upsilon}{\gammai}\right)^{1-2\chis}\right)\\
        B &= \dfrac{-A}{4\chis} \left[ 2A + \dfrac{1}{2\chis+1-3\nu} + \dfrac{1}{2-2\chis}+ \dfrac{1}{1-2\chis} +\ln\left(\dfrac{\Upsilon}{\gammai}\right) \right]
    \end{align}

\end{document}